\documentclass[letterpaper, submit]{AAS}			
\usepackage{booktabs}
\usepackage{bm}
\usepackage{amsmath}
\usepackage[colorlinks=true, pdfstartview=FitV, linkcolor=black, citecolor= black, urlcolor= black]{hyperref}
\usepackage{overcite}
\usepackage{footnpag}			      	

\usepackage{amssymb}
\usepackage{siunitx}
\usepackage{graphicx}
\usepackage{subcaption}
\usepackage{tikz}
\usetikzlibrary{calc}
\usepackage{tcolorbox}
\usepackage{standalone}

\PaperNumber{25-XXX}

\newcommand{\orcid}[1]{\href{https://orcid.org/#1}{\includegraphics[scale=.012]{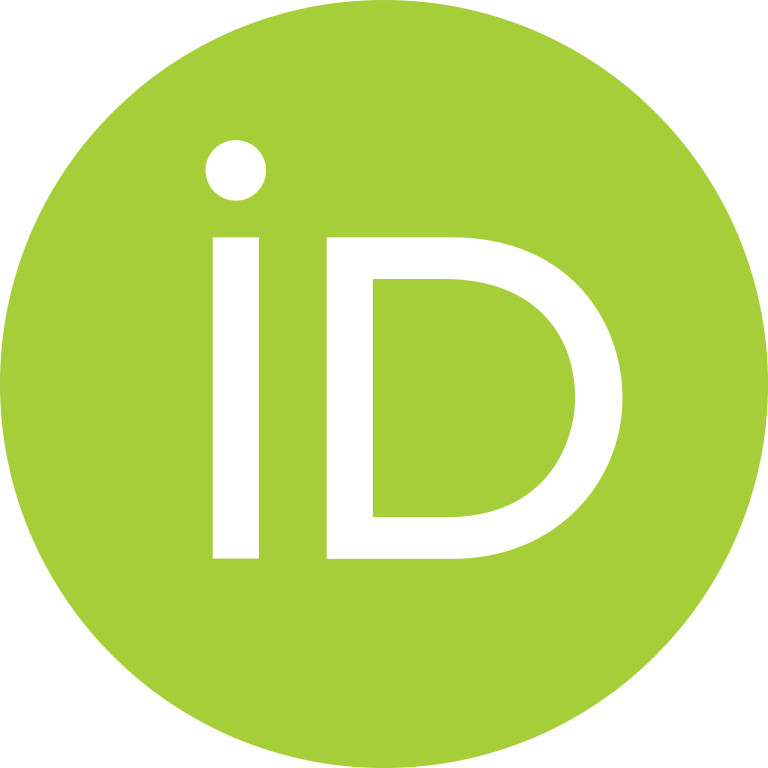}}}

\begin{document}

\title{
Global Search for Optimal Low Thrust Spacecraft Trajectories using Diffusion Models and the Indirect Method
\thanks{
An abbreviated version of this paper has been accepted for presentation at the 35th AAS/AIAA Space Flight Mechanics Meeting, Kaua'i Hawaii, Jan. 2025, as Paper 25-319 with title \emph{Global Search for Optimal Low Thrust Spacecraft Trajectories using Diffusion Models and the Indirect Method}.
}
}

\author{Jannik Graebner \orcid{https://orcid.org/0009-0003-2497-124X} \thanks{PhD Student, Department of Mechanical and Aerospace Engineering, Princeton University, NJ, USA.}
\ and Ryne Beeson \orcid{0000-0003-2176-0976} \thanks{Assistant Professor, Department of Mechanical and Aerospace Engineering, Princeton University, NJ, USA.}
}

\maketitle{} 		

\begin{abstract}
The global search for optimal long time-duration, low-thrust spacecraft trajectories is a computationally expensive problem, that is characterized by clustering patterns in locally optimal solutions.
During preliminary mission design, mission parameters have not been fully defined yet, necessitating that trajectory designers efficiently generate high-quality control solutions across a wide range of different scenarios.
Generative machine learning models can be trained to learn how the solution structure varies with respect to an evolving mission parameter, thereby accelerating the global search for a large number of missions with varying parameters.
In this work, state-of-the-art diffusion models are integrated with the indirect approach for trajectory optimization within a global search framework.
The main difficulty with the indirect method lies in generating good initial guesses for the non-intuitive costate variables, which are crucial for solver convergence. 
By training a diffusion model to learn the structure of high-quality solutions in costate space, it can generate initial costate guesses for new mission parameters.
This framework is tested on two low-thrust transfers of different complexity in the circular restricted three-body problem.
By generating and analyzing a training data set, we develop mathematical relations and techniques to understand the complex structures in the costate domain of locally optimal solutions for these problems.
A diffusion model is trained on this data and successfully predicts how the costate solution structure changes, based on the maximum spacecraft thrust magnitude.
We warm-start a numerical solver with initial costates  sampled from the diffusion model for problems with unseen thrust magnitudes and compare the number of solutions generated per minute to samples from a uniform distribution and from an adjoint control transformation.
Results show that the diffusion model accelerates the global search process by one to two orders of magnitude, allowing for rapid generation of high-quality solutions.

\end{abstract}

\section{Introduction}
Low-thrust trajectory optimization is a global search problem with multiple objectives and constraints. 
Finding optimal solutions while minimizing fuel consumption and time of flight is a complex, computationally expensive and time-consuming endeavor tackled in the preliminary mission design phase.
This task is further complicated by key mission parameters and objectives not being fixed yet.
The flight-dynamics team therefore surveys a broad design space, generating hundreds to thousands of candidate trajectories to understand how performance varies across plausible parameter sets.
For example, varying the maximum thrust magnitude can result in families of optimal trajectories with fundamentally different characteristics.
Adjustments might also be required mid-mission, as exemplified by the BepiColombo spacecraft, operated by ESA and JAXA.
In May 2024, reduced solar cell performance limited the thrusters’ operation to $90\%$ of maximum power \cite{EuropeanSpaceAgency.2024}. 
Ideally, control solutions for such new constraints would be generated swiftly, either by ground engineers or autonomous on-board algorithms. 
We present a framework that uses generative machine learning to speed up the global search for trajectory optimization problems with varying mission parameters.

Previous studies have demonstrated that locally optimal solutions to low-thrust trajectory optimization problems tend to form clusters \cite{Li.872023,beeson2024globalsearchoptimalspacecraft,Vasile.2006,doi:10.2514/6.2012-4517,Yam.2011}. 
Global search algorithms like monotonic basin hopping (MBH) aim to exploit this clustering to improve efficiency in generating solutions \cite{Wales.1997}. 
However, sampling distributions for MBH are typically chosen for their simplicity such as uniform, Gaussian, Cauchy, or Pareto distributions \cite{englander2014tuning}. 
Their parameters often require manual tuning, although there has been work on adaptive self-tuning variants of MBH \cite{Englander2021}.
Li et al. \cite{Li.872023} proposed an amortized global search framework that learns a conditional probability distribution supported on local basins of attraction for high-quality solutions using deep generative models. 
Their distribution was conditioned on the spacecraft’s maximum thrust, initially employing a conditional variational autoencoder (CVAE), later enhanced with a long short-term memory (LSTM), alongside a numerical solver to generate solutions for new thrust values. 
Beeson et al. \cite{beeson2024globalsearchoptimalspacecraft} built on this work by mathematically defining the parameterized global search problem as the task of sampling a conditional
probability distribution, which focuses on basins of attractions of high-quality solutions. 
Subsequent studies \cite{Li.2222024, graebner2024learningoptimalcontroldynamical} found that diffusion models (DM) outperformed the CVAE + LSTM approach for trajectory optimization. 
Diffusion probabilistic models, are state-of-the-art generative models, that have recently captured widespread interest, due to their strong performance in conditional image and video generation \cite{Dhariwal.5112021}.

The novel contribution of this work is the integration of a diffusion model framework with indirect optimal control for spacecraft trajectory optimization. 
While this builds on the work of Li et al. \cite{Li.872023,Li.2222024}, their research has focused solely on direct optimal control methods. Direct methods use transcription techniques to approximate dynamics and control, reformulating the optimal control problem as a nonlinear program. 
In contrast, indirect methods derive the first-order necessary conditions for optimality, where the additional costate variables are introduced. 
This results in a two-point boundary value problem. 
Although direct methods are now the standard in spacecraft trajectory optimization due to their simpler setup and robust convergence properties, especially with poor initial guesses, indirect methods provide a lower-dimensional solution space where each solution satisfies the first-order optimality conditions \cite{Betts.1998}. 
The main difficulty with the indirect method lies in generating good initial guesses for the non-intuitive costate variables, which are crucial for solver convergence. 
This challenge drives the use of generative learning in this context. 
By training a diffusion model to learn the structure of high-quality solutions in costate space, it can generate initial costate guesses for new conditional parameters. 
In general we use the expression conditional parameter to describe a physical parameter of the problem definition that, when varied, results in different families of trajectories.

Several studies have addressed the challenge of estimating initial costates using machine learning techniques.
Parrish and Scheeres employed a feedforward neural network to correct costate initializations based on state constraint violations \cite{Parrish.2018b}. 
Similarly, Shi et al. trained a neural network to provide accurate initial costates for warm-starting the optimization algorithm \cite{Shi.2022}. 
However, their work primarily focused on simplified dynamics, with the aim of implementing optimal control in real time.
Tang and Hauser \cite{Tang.2019} adopted a data-driven approach, determining initial costate guesses for new problems by applying a nearest-neighbor algorithm to precomputed solutions.
Yin et al. combined a deep neural network with a numerical solver to efficiently generate optimal low-thrust trajectories using the indirect method, though their research was limited to simple two-body dynamics \cite{Yin.2020}. 
Although the use of machine learning to approximate initial costates is not a new concept, it has yet to be extended to solution generation for long duration low-thrust transfers in complex multibody dynamics.

While the diffusion model facilitates the rapid generation of solutions for new conditional parameters, it requires a substantial amount of data for training. 
A key contribution of this work is the implementation of efficient methods to generate this training data. 
This is achieved by integrating the astrodynamics software \texttt{pydylan} \cite{Beeson.Aug.2022} with adjoint control transformations (ACTs) and a modified version of the preliminary screening algorithm introduced by Russell \cite{Russell.2007}. 
The generative learning framework’s capabilities are demonstrated through two low-thrust transfers within the circular restricted three-body problem (CR3BP), with both transfers conditioned on the spacecraft’s maximum thrust. 
Training data is generated for fixed values of this parameter, and the diffusion model is then tested by generating solutions for a new thrust level not included in the training set. Results show that the diffusion model accelerates the global search process by one to two orders of magnitude, allowing for rapid generation of high-quality solutions.



\section{Problem Formulation}
\subsection{Optimal Control Problem}
The goal of optimal control theory is to determine a control signal which causes a process to satisfy a set of physical constraints, while maximizing or minimizing a performance metric \cite{Kirk.2004}. 
This performance index is formulated as a cost or objective functional $J$, including a terminal cost $\phi$ and a running cost $\mathcal{L}$. A generic formulation of the problem is given by:
\begin{align}
    \label{equation: cost functional}
    \min \left\{J(\boldsymbol{u}) \equiv \phi(\boldsymbol{x}(t_f), t_f) + \int_{t_0}^{t_f} \mathcal{L}(\boldsymbol{x}(t), \boldsymbol{u}(t), t) dt, \quad 
    \textrm{subj. to satisfaction of Eqs. } 
    \eqref{equation: evolution differential equation}, \eqref{equation: initial and final boundary conditions}, \eqref{equation: path constraints}
    \right\}, 
\end{align}
where $\boldsymbol{u}(t)$ is the control, 
and $\boldsymbol{x}(t)$ is the state of the system. 
Time is represented by $t\in [t_0,t_f]$, with $t_0$ denoting the initial time and $t_f$ the final time. 
The state evolves according to
\begin{align}
    \label{equation: evolution differential equation}
    \dot{\boldsymbol{x}}(t) = \boldsymbol{f}(\boldsymbol{x}(t), \boldsymbol{u}(t), t), \quad \forall t \in [t_0, t_f],
\end{align}
and satisfies the initial and terminal boundary conditions, 
\begin{align}
    \label{equation: initial and final boundary conditions}
    \boldsymbol{x}(t_0) = \boldsymbol{x}_0, \quad  \boldsymbol{\psi}\left( \boldsymbol{x}(t_f), t_f \right) = \boldsymbol{0},
\end{align}
as well as the equality path constraints, 
\begin{align}
    \label{equation: path constraints}
    \boldsymbol{\xi}(\boldsymbol{x}(t), \boldsymbol{u}(t), t) = \boldsymbol{0}, \quad & \forall t \in [t_0, t_f].
\end{align}
The system’s dynamics are represented by the vector field $\boldsymbol{f}$, which encompasses both the inherent natural dynamics and any perturbations, including those induced by the control.
The initial state is prescribed as the vector $\boldsymbol{x}_0$, while terminal boundary conditions are defined through the function $\boldsymbol{\psi}$.

\subsection{Low-thrust Spacecraft Trajectory Optimization}

Optimizing a spacecraft’s trajectory is a complex, nonlinear optimal control problem. 
The goal is to determine a trajectory that satisfies the specified initial and terminal conditions for the mission while minimizing a performance measure \cite{Conway.2010}.
A common objective is to minimize the amount of propellant required for the mission, which maximizes the fraction of the total mass available for other systems. 
The objective function in this work is formulated as:
\begin{equation}
	\label{eq:cost function}
	J = -km_f,
\end{equation}
where $m_f$ is the final mass of the spacecraft and $k$ is a constant multiplier with $k>0$.
Additionally, an upper bound is imposed on the flight time to prevent the optimizer from excessively trading time for propellant mass.
We consider a low-thrust engine with constant specific impulse $I_{sp}$ and exhaust velocity $c=I_{sp}g_0$, where $g_0$ represents standard gravity.
The state variable $\boldsymbol{x}$, which comprises the spacecraft's position $\boldsymbol{r}\in \mathbb{R}^{3}$,  velocity $\boldsymbol{v}\in \mathbb{R}^{3}$ and mass $m\in \mathbb{R}$, and the dynamics of the system are given by:
\begin{equation}
\label{Equation: dynamics}
\boldsymbol{x}=\begin{pmatrix}
	\boldsymbol{r} \\
	\boldsymbol{v} \\
	m
	\end{pmatrix},
	\in \mathbb{R}^{7} \quad \text{and} \quad
    \dot{\boldsymbol{x}} = \boldsymbol{f}(\boldsymbol{x},\boldsymbol{u}) =
	\begin{pmatrix}
	\dot{\boldsymbol{r}} \\
	\dot{\boldsymbol{v}} \\
	\dot{m}
	\end{pmatrix}
	=
	\begin{pmatrix}
	\boldsymbol{v} \\
	\boldsymbol{g}(\boldsymbol{r},\boldsymbol{v}) + \frac{T}{m}\hat{\boldsymbol{u}} \\
	-\frac{T}{c}
	\end{pmatrix},
\end{equation}
where $\boldsymbol{g}(\boldsymbol{r},\boldsymbol{v})$ represents a vector field dependent on the position $\boldsymbol{r}$ and velocity $\boldsymbol{v}$.
The thrust magnitude $T$ is controlled by a throttle variable $\sigma \in [0,1]$ with $T = \sigma T_{max}$, where $T_{max}$ is the engine’s maximum thrust. 
To enforce throttle bounds as an equality constraint, a slack control variable $\zeta$ is introduced, with $\sigma = \sin^2 \zeta$. 
The thrust direction is described by a unit vector $\hat{\boldsymbol{u}}\in \mathbb{R}^{3}$.
The three control variables, $\hat{\boldsymbol{u}}$, $T$ and $\zeta$, are summarized in the control vector $\boldsymbol{u}=(\hat{\boldsymbol{u}}^T,T,\zeta)^T$.
Two equality path constraints are employed to enforce the bounds on the control variables:
\begin{equation}
    \boldsymbol{\xi}(\boldsymbol{x}(t), \boldsymbol{u}(t), t) = 
    \begin{pmatrix}
	\hat{\boldsymbol{u}}^T\hat{\boldsymbol{u}} - 1 \\
    T-T_{max}\sin^2\zeta
    \end{pmatrix} = \boldsymbol{0}.
\end{equation}
The initial and terminal boundary conditions are:
\begin{equation}
\label{Equation: Boundary conditions trajectory}
	\boldsymbol{x}(t_0) = \boldsymbol{x}_0, \quad\boldsymbol{\Psi}(\boldsymbol{x}(t_f)) =
    \begin{pmatrix}
	\boldsymbol{r}(t_f)- \boldsymbol{r}_f\\
    \boldsymbol{v}(t_f)- \boldsymbol{v}_f
    \end{pmatrix} = \boldsymbol{0},
\end{equation}
where the initial state is specified by the user as $\boldsymbol{x}_0$ and the terminal position and velocity are prescribed as $\boldsymbol{r}_f$ and $\boldsymbol{v}_f$.
The final time is a free parameter that is determined by the optimizer.
\subsection{Indirect Approach}
The indirect method applies Pontryagin’s Minimum Principle \cite{Liberzon.2012} to derive the necessary conditions for optimality. 
The control Hamiltonian of the problem, which we will just refer to as the Hamiltonian for brevity, is expressed as 
\begin{equation}
	\label{Equation: Problem Hamiltonian}
	H = \mathcal{L} + \boldsymbol{\lambda}^T \boldsymbol{f}(\boldsymbol{x},\boldsymbol{u}) = \boldsymbol{\lambda}_r^T \boldsymbol{v} + \boldsymbol{\lambda}_v^T (\boldsymbol{g}(\boldsymbol{r},\boldsymbol{v}) + \frac{T}{m}\hat{\boldsymbol{u}}) - \lambda_m\frac{T}{c},
\end{equation}
where the costate vector $\boldsymbol{\lambda} =(\boldsymbol{\lambda}_r^T,\boldsymbol{\lambda}_v^T, \lambda_m)^T \in \mathbb{R}^{7}$ includes the position costate vector $\boldsymbol{\lambda}_r\in \mathbb{R}^{3}$, the velocity costate vector $\boldsymbol{\lambda}_v\in \mathbb{R}^{3}$ and the mass costate $\lambda_m\in \mathbb{R}$.
According to Pontryagin’s Minimum Principle \cite{Pontryagin.2018}, the Hamiltonian in Eq. \eqref{Equation: Problem Hamiltonian} needs to be minimized, resulting in $\hat{\boldsymbol{u}} = -\boldsymbol{\lambda}_v/\lambda_v$. 
Substituting this expression into Eq. \eqref{Equation: Problem Hamiltonian} yields:
\begin{equation}
	\label{Equation: Hamiltonian without control}
	H = \boldsymbol{\lambda}_r^T \boldsymbol{v} + \boldsymbol{\lambda}_v^T \boldsymbol{g}(\boldsymbol{r},\boldsymbol{v}) - S\frac{T}{m},\quad S = \lambda_v + \lambda_m m / c,
\end{equation}
where $S$ is the switching function that determines $T$ to satisfy the minimum principle, resulting in “bang-bang” behavior \cite{Conway.2010} summarized in the control law:
\begin{equation}
	\label{Equation: Control Law}
	\hat{\boldsymbol{u}} = -\frac{\boldsymbol{\lambda_v}}{\lambda_v}, \quad
	\sigma = 
	\begin{cases}
	0 & \text{if } S < 0 \\
	1 & \text{if } S > 0 \\
	0 \leq \sigma \leq 1 & \text{if } S = 0,
	\end{cases}
\end{equation}
which allows the control variables to be determined for given costates.
The equations of motion for the costates are derived using the necessary conditions for optimality \cite{Liberzon.2012}:
\begin{equation}
\label{Equation: Costate equation of motion}
\dot{\boldsymbol{\lambda}} = \begin{pmatrix}
\dot{\boldsymbol{\lambda}}_r \\
\dot{\boldsymbol{\lambda}}_v \\
\dot{\lambda}_m
\end{pmatrix} = \begin{pmatrix}
- \boldsymbol{G}^T \boldsymbol{\lambda}_v \\
-\boldsymbol{\lambda}_r - \boldsymbol{H}^T \boldsymbol{\lambda}_v \\
-\lambda_v T / m^2
\end{pmatrix}, \quad\text{where}\quad
\mathbf{G} = \frac{\partial \mathbf{g}}{\partial \mathbf{r}} \quad \text{and} \quad \mathbf{H} = \frac{\partial \mathbf{g}}{\partial \mathbf{v}}.
\end{equation}
Combining the equations of motion for states and costates yields  a two point boundary value problem subject to the transversality conditions: 
\begin{equation}
    \label{Equation: lam_m=-k}
    \lambda_m(t_f)=-k \quad \text{and} \quad H(\boldsymbol{x}(t_f), \boldsymbol{u}(t_f), \boldsymbol{\lambda}(t_f), t_f)=0.
\end{equation}
Since the factor $k>0$ can be chosen freely, it provides an additional degree of freedom.
The core concept of the indirect method is to discretize and numerically solve this problem.

\subsection{Circular Restricted Three Body Problem}
Selecting a suitable dynamical model is vital to space mission design, as each strikes a different balance between realism and complexity. 
The circular restricted three-body problem (CR3BP) is commonly used during preliminary mission design for multi-body missions, as it simplifies the problem while utilizing dynamical structures for effective mission planning.
In the CR3BP, a primary celestial body of mass $m_1$ and a secondary celestial body of mass $m_2$ ($m_1>m_2$) move in circular orbits around their common center of mass.
The CR3BP describes the motion of a third body of negligible mass, that does not affect the orbits of the larger bodies \cite{Szebehely.2012}.
Variables in the CR3BP are expressed using a system of normalized units, referred to as natural units (NU): the unit of mass is $m_1 + m_2$, the distance unit (DU) is the constant distance between primary and secondary, and the time unit (TU) is chosen such that their orbital period around the barycenter is $2\pi$. 
The system’s only parameter is the mass ratio $\mu = \frac{m_2}{m_1 + m_2}$.
The third body’s motion is described in a rotating reference frame, where the primary and secondary are fixed along the  $r_1$ -axis at  $(-\mu, 0, 0)$  and  $(1 - \mu, 0, 0)$, with their barycenter at the origin. The  $r_3$ -axis points in the direction of angular momentum, while the  $r_2$-axis completes the right-handed system. 
The motion of the third body under the gravitational influence of $m_1$ and $m_2$ follows the dynamical system:
\begin{equation}
    \label{Equation: CR3BP dynamical system}
    \ddot{\boldsymbol{r}} = 
    \begin{pmatrix}
        \ddot{r}_1 \\
        \ddot{r}_2 \\
        \ddot{r}_3 
    \end{pmatrix}
    = \boldsymbol{g}(\boldsymbol{r},\boldsymbol{v}) =
    \begin{pmatrix}
        2v_2 + r_1 - (1-\mu)\frac{r_1+\mu}{\rho_1^3} - \mu\frac{r_1-1+\mu}{\rho_2^3} \\
        -2v_1 + r_2 - (1-\mu)\frac{r_2}{\rho_1^3} - \mu\frac{r_2}{\rho_2^3} \\
        -(1-\mu)\frac{r_3}{\rho_1^3} - \mu\frac{r_3}{\rho_2^3}
    \end{pmatrix},
\end{equation}
where $\rho_1= \sqrt{(r_1+\mu)^2+r_2^2+r_3^2}$ and $\rho_2=\sqrt{(r_1-1+\mu)^2+r_2^2+r_3^2}$ describe the distance to the primary and secondary.
Koon et al. provide a detailed derivation of the equations of motion for the CR3BP \cite{Koon.}.

\section{Methodology}
\subsection{Adjoint Control Transformation}
\label{sec: ACT}
When using an indirect method, an initial guess for the costate variables is required to solve the boundary value problem. 
This is challenging due to the small convergence radius and lack of physical intuition for costates \cite{Dixon.1972}. 
During data generation we employ a method referred to as adjoint control transformation (ACT) to generate initial costate guesses\cite{Dixon.1981}. The ACT method maps a set of physical control variables to the non-physical costates: $\mathcal{M}: (\varphi, \dot{\varphi}, \beta, \dot{\beta}, S, \dot{S}) \rightarrow (\boldsymbol{\lambda}_{r}, \boldsymbol{\lambda}_{v})$, where $\varphi$ and $\beta$ represent the in-plane and out-of-plane angles of the thrust vector in a velocity-centered spacecraft coordinate frame.
We evaluate this mapping at $t_0$ by uniformly sampling initial guesses for $\varphi_0$, $\beta_0$, $\Dot{\varphi}_0$, $\Dot{\beta}_0$, $S_0$ and $\Dot{S}_0$ within problem-specific ranges, yielding reasonable initial guesses for $\boldsymbol{\lambda}_{r,0}$ and $\boldsymbol{\lambda}_{v,0}$.
No mapping is needed for the mass costate at $t_0$, as we exploit the degree of freedom introduced by the factor $k$ by setting $\lambda_{m,0}=-1$.
The transversality condition $\lambda_m(t_f)=-k$ is satisfied, since $\dot{\lambda}_m\le0$.
\begin{figure}[b!]
	\centering
    \resizebox{0.4\textwidth}{!}{\normalsize

\begin{tikzpicture}
    \pgfmathsetmacro{\Px}{3.9}
    \pgfmathsetmacro{\Py}{3.7}
    \coordinate (P) at (\Px, \Py);
    
    \pgfmathsetmacro{\angleX}{100}
    \pgfmathsetmacro{\angleY}{190}
    \pgfmathsetmacro{\angleZ}{225}
    
    \pgfmathsetmacro{\lengthXY}{1.68}
    \pgfmathsetmacro{\lengthZ}{0.7}
    
    \fill (P) circle[radius=0.8pt];
    
    \draw[->, semithick,>=stealth] (P) -- ++(\angleX:\lengthXY) node[right] {$\hat{\boldsymbol{v}}$};
    \draw[->, semithick,>=stealth] (P) -- ++(\angleY:\lengthXY) node[above] {$\hat{\boldsymbol{w}}$};
    \draw[->, semithick,>=stealth] (P) -- ++(\angleZ:\lengthZ) node[left] {$\hat{\boldsymbol{h}}$};

    \draw[->, semithick,  teal,>=stealth] (P) -- ++(\angleX:{1.6*\lengthXY}) node[ right] {$\boldsymbol{v}$};
    \pgfmathsetmacro{\angleU}{-55} 
    \pgfmathsetmacro{\lengthU}{2.5} 
    \draw[->, semithick, red,>=stealth] (P) -- ++(\angleU:\lengthU) coordinate (U)  node[below] {$\boldsymbol{u}$};

    \pgfmathsetmacro{\angleW}{-40} 
    \pgfmathsetmacro{\lengthW}{2.0} 
    \draw[->, semithick, dotted,>=stealth] (P) -- ++(\angleW:\lengthW) coordinate (W);

    \draw ($(W)!0.1!(P)$) -- ++(\angleW-60:0.2) coordinate (A) -- ++(\angleW:0.2) coordinate (B);
    \draw (A) -- (B);

    \draw[->, semithick, dotted, >=stealth] (W) -- (U) node[below] {};

    \draw[->, orange, >=stealth] ($(P)!0.8!(W)$) arc[start angle=\angleW, end angle=\angleU, radius=1.6] node[pos = 0.0, above] {$\beta$};

    \begin{scope}
    \fill[orange, opacity=0.3] (P) -- ++(\angleW:1.6cm) arc[start angle=\angleW, end angle=\angleU, radius=1.6cm] -- ++(\angleU-180:0.5cm) -- cycle;
    \end{scope}

    \pgfmathsetmacro{\angleV}{\angleX}
    \pgfmathsetmacro{\deltaVW}{\angleW-\angleV}
    \draw[->, blue, >=stealth] (P) +(\angleV:{0.3*\lengthXY}) arc[start angle=\angleV, delta angle=(360+\deltaVW), radius=0.3*\lengthXY] node[pos=0.3, right] {$\varphi$};

    \begin{scope}
    \fill[blue, opacity=0.3] (P) -- ++(\angleV:0.3*\lengthXY) arc[start angle=\angleV, delta angle=(360+\deltaVW), radius=0.3*\lengthXY]; 
    \end{scope}

    \draw[->, >=stealth, thick] (-1.4, 0) -- (8.4, 0) node[below right] {$r_1$};
    \draw[->, >=stealth, thick] (0, -1.12) -- (0, 6.72) node[above left] {$r_2$};
    \draw[->, >=stealth, thick] (0.784*1.4142, 0.784*1.4142) -- (-0.784*1.4142, -0.784*1.4142) node[below right] {$r_3$};

    \path (-1.4,-1.12) rectangle (6.72,5.6);

    \fill[brown] (-1.12, 0) circle (5pt);
    \node at (-1.12, 0.28) {$m_1$};
    \fill[brown] (6.72, 0) circle (3pt);
    \node at (6.72, 0.28) {$m_2$};
\end{tikzpicture}}
    \caption{Display of the spacecraft-centered coordinate frame relative to the CR3BP frame. The thrust vector $\boldsymbol{u}$ is shown, along with the in-plane and out-of-plane thrust angles $\varphi$ and $\beta$.}
	\label{fig:Spacecraft COS}
\end{figure}
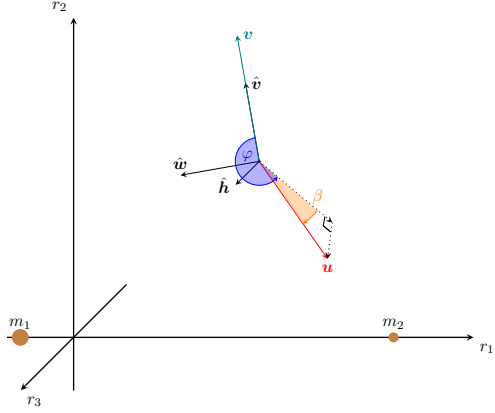

A spacecraft centered coordinate frame and its transformation to the CR3BP frame are defined according to the illustration in Fig. \ref{fig:Spacecraft COS}. 
The x-axis, denoted by $\hat{\boldsymbol{v}}$, is aligned with the spacecraft’s velocity vector, the z-axis, denoted by $\hat{\boldsymbol{h}}$, points in the direction of the spacecraft’s angular momentum $\boldsymbol{h} = \boldsymbol{r} \times \boldsymbol{v}$ and the y-axis $\hat{\boldsymbol{w}}$ completes the right-handed coordinate system:
\begin{equation}
    \hat{\boldsymbol{v}} = \frac{\boldsymbol{v}}{v}, \quad \hat{\boldsymbol{h}} = \frac{\boldsymbol{h}}{h} = \frac{\boldsymbol{r} \times \boldsymbol{v}}{\|\boldsymbol{r} \times \boldsymbol{v}\|}, \quad \hat{\boldsymbol{w}} = \hat{\boldsymbol{h}} \times \hat{\boldsymbol{v}}.
\end{equation}
The derivative of the vectors $\hat{\boldsymbol{v}}$, $\hat{\boldsymbol{h}}$ and $\hat{\boldsymbol{w}}$ are:
\begin{equation}
    \dot{\hat{\boldsymbol{v}}} = \frac{\dot{\boldsymbol{v}}}{v} - \frac{\boldsymbol{v} \dot{v}}{v^2}, \quad \dot{\hat{\boldsymbol{h}}} = \frac{\dot{\boldsymbol{h}}}{h} - \frac{\boldsymbol{h} \dot{h}}{h^2}
    , \quad \dot{\hat{\boldsymbol{w}}} = \dot{\hat{\boldsymbol{h}}} \times \hat{\boldsymbol{v}} + \hat{\boldsymbol{h}} \times \dot{\hat{\boldsymbol{v}}},
\end{equation}
where $\dot{v} = v \cdot \dot{\boldsymbol{v}}/v
$, $\dot{\boldsymbol{h}} = \boldsymbol{r} \times \dot{\boldsymbol{v}}$ and $\dot{h} = h \cdot \dot{\boldsymbol{h}}/h$.
Using these vectors, the transformation matrix $\boldsymbol{R}$, which maps a vector from the spacecraft coordinate frame to the CR3BP frame, along with its derivative $\dot{\boldsymbol{R}}$, are defined as $\boldsymbol{R} = [ \hat{\boldsymbol{v}} \, | \, \hat{\boldsymbol{w}} \,| \, \hat{\boldsymbol{h}} ]$ and $\dot{\boldsymbol{R}} = [ \dot{\hat{\boldsymbol{v}}}\, | \,\dot{\hat{\boldsymbol{w}}} \, | \, \dot{\hat{\boldsymbol{h}}} ]$.
The thrust angles are used to express the thrust direction unit vector and its derivative in the new coordinate frame (denoted by a prime symbol):
\begin{align}
    \hat{\boldsymbol{u}}' = 
    \begin{bmatrix}
    \cos(\varphi) \cos(\beta) \\
    \sin(\varphi) \cos(\beta) \\
    \sin(\beta)
    \end{bmatrix}, \quad
    \dot{\hat{\boldsymbol{u}}}' = 
    \begin{bmatrix}
    -\sin(\varphi) \dot{\varphi} \cos(\beta) - \cos(\varphi) \sin(\beta) \dot{\beta} \\
    \cos(\varphi) \dot{\varphi} \cos(\beta) - \sin(\varphi) \sin(\beta) \dot{\beta} \\
    \cos(\beta) \dot{\beta}
    \end{bmatrix}.
\end{align}
Using the transformation matrices, $\hat{\boldsymbol{u}}$ and $\dot{\hat{\boldsymbol{u}}}$ are expressed in the CR3BP coordinate frame:
\begin{align}
    \label{eq: thrust transformation}
    \hat{\boldsymbol{u}} = \boldsymbol{R} \hat{\boldsymbol{u}}', \quad
    \dot{\hat{\boldsymbol{u}}} = \dot{\boldsymbol{R}} \hat{\boldsymbol{u}}' + \boldsymbol{R} \dot{\hat{\boldsymbol{u}}}'.
\end{align}
The mapping is given by first evaluating:
\begin{align}
    \label{Equation: ACT1}
    \lambda_{\boldsymbol{v}} = S - \lambda_{m} \frac{m}{c}, \quad \dot{\lambda}_{\boldsymbol{v}} = \dot{S} - \lambda_{m} \frac{\dot{m}}{c} - \dot{\lambda}_{m} \frac{m}{c},
\end{align}
and then:
\begin{align}
    \label{Equation: ACT2}
    \boldsymbol{\lambda_v} = -\lambda_v \hat{\boldsymbol{u}}, \quad
    \dot{\boldsymbol{\lambda}}_v = -\dot{\lambda}_v \hat{\boldsymbol{u}} - \lambda_v \dot{\hat{\boldsymbol{u}}}, \quad
    \boldsymbol{\lambda}_r = -\dot{\boldsymbol{\lambda}}_v - \boldsymbol{H}^T \boldsymbol{\lambda}_v.
\end{align}

\subsection{Global Search Strategy}
\label{section: global search strategies}
In this work, \texttt{pydylan}, the Python interface of the astrodynamics software package Dynamically Leveraged (N) Multibody Trajectory Optimization (DyLAN) \cite{Beeson.Aug.2022}, is used for trajectory optimization. 
DyLAN integrates dynamical systems techniques with automated global optimization algorithms and uses the gradient-based sequential quadratic programming software SNOPT (Sparse Nonlinear OPTimizer) \cite{Gill.2005} to solve nonlinear programs. 
For the indirect method, the decision variable provided to SNOPT is defined as:
\begin{equation}
    \label{Equation: decision vector indirect}
    \overline{\boldsymbol{u}} = (\tau_s,\tau_i,\tau_f,\boldsymbol{\lambda}_{r,0}^T,\boldsymbol{\lambda}_{v,0}^T,{\lambda}_{m,0})^T,
\end{equation}
where $\overline{\boldsymbol{u}}$ is a $10$-dimensional vector. 
The trajectory maneuver occurs during a shooting phase of duration $\tau_s$, which is enclosed by an initial and final coasting phase of duration $\tau_i$ and $\tau_f$. 
In this work, we set $\tau_i=0$ and treat $\tau_s$ and $\tau_f$ as free parameters in the optimization process, although upper and lower bounds are enforced. 
\begin{figure}[b!]
	\centering
	\includegraphics[width=0.9\textwidth]{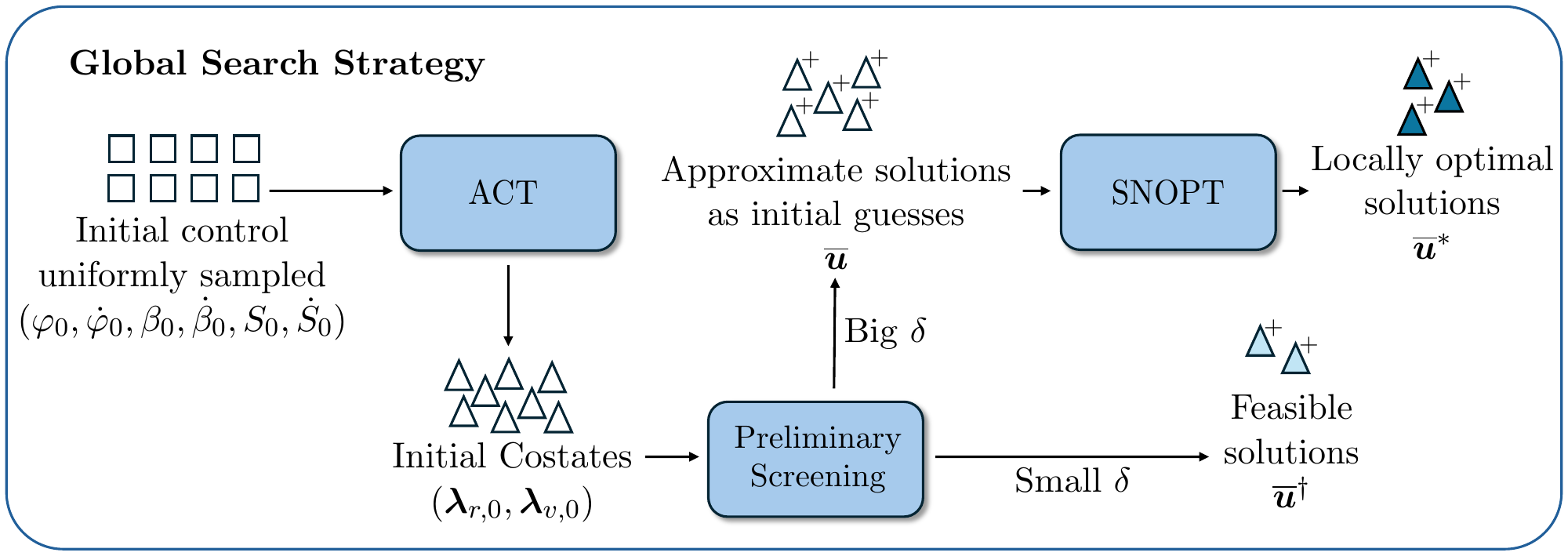}
	\caption{Visualization of the global search strategy used for data generation. Two different approaches are shown: the lower path, where the preliminary screening approach is directly used to generate feasible solutions, and the upper path where it is combined with SNOPT.}
	\label{fig: global search}
\end{figure}

Since SNOPT is a local, gradient-based optimizer, we pair it with a global search strategy to explore the entire search space and generate diverse locally optimal solutions. 
This strategy, outlined in Fig. \ref{fig: global search}, integrates the ACT method with a preliminary screening algorithm based on work from Russell \cite{Russell.2007}.
The preliminary screening algorithm quickly assesses the quality of an initial costate vector by propagating it forward in time for a maximum shooting time $\tau_{s,max}$.
The detailed steps of the algorithm are presented as pseudo-code in Fig. \ref{fig: preliminary screening}.
\begin{figure}[t!]
\centering
\begin{tcolorbox}[colframe=black, colback=white, boxrule=0.5mm, width=\textwidth]
\begin{enumerate}
    \item Set fixed values: $\boldsymbol{r}(0)=\boldsymbol{r}_0$, $\boldsymbol{v}(0)=\boldsymbol{v}_0$, $m(0)=m_0$, $\lambda_{m,0}=-1$, $\tau_i=0$
    \item Generate k-d tree for states on target orbit or manifold 
    \item \textbf{DO WHILE} ($j \leq j_{max}$ with $j\in \mathbb{N}$)
    \begin{enumerate}
        \item $j = j + 1$
        \item Apply ACT method to obtain $\boldsymbol{\lambda}_{r,0}$ and $\boldsymbol{\lambda}_{v,0}$
        \item Propagate spacecraft states and costates according to Eqs. \eqref{Equation: dynamics}, \eqref{Equation: Control Law} and \eqref{Equation: Costate equation of motion}  for $\tau_{s,max}$ and save $\boldsymbol{r}(t)$ and $\boldsymbol{v}(t)$
        \item $\delta_{c,min}=\infty$
        \item \textbf{FOR} $t=0$ to $t=\tau_{s,max}$ \textbf{DO}
        \begin{enumerate}
            \item Query k-d tree to perform nearest neighbor search with 
            $\begin{pmatrix}
            \boldsymbol{r}(t) \\
            \boldsymbol{v}(t)
            \end{pmatrix}$
            \item k-d tree returns closest point on target orbit or manifold 
            $\begin{pmatrix}
            \boldsymbol{r}_f(\tau_f^*) \\
            \boldsymbol{v}_f(\tau_f^*)
            \end{pmatrix}$ 
            with corresponding final coast time $\tau_f^*$
            \item Calculate maximum constraint violation $\delta_c=\left\Vert\begin{pmatrix}
            \boldsymbol{r}(t) - \boldsymbol{r}_f(\tau_f^*) \\
            \boldsymbol{v}(t) - \boldsymbol{v}_f(\tau_f^*)
            \end{pmatrix}\right\Vert_\infty$
            \item \textbf{IF $\delta_c < \delta_{c,min}$} 
            \begin{enumerate}
                \item Update best values: $\delta_{c,min}=\delta_c$, $\tau_s=t$, $\tau_f = \tau_f^*$
            \end{enumerate}
        \end{enumerate}
        \item \textbf{END FOR}
        \item \textbf{IF} $\delta_{c,min}<\delta$
            \begin{enumerate}
                \item Switch to local optimization routine with initial guess: \\ $\overline{\boldsymbol{u}}=(\tau_s,\tau_i,\tau_f,\boldsymbol{\lambda}_{r,0}^T,\boldsymbol{\lambda}_{v,0}^T,\lambda_{m,0})$
            \end{enumerate}
        \end{enumerate}
    \end{enumerate}
\end{tcolorbox}
\caption{Preliminary screening algorithm.}
\label{fig: preliminary screening}
\end{figure}

First, the initial state and other fixed variables are set. The terminal boundary condition is defined to either be a target orbit or stable invariant manifold of an orbit. 
This provides an additional degree of freedom, described by the final coast time $\tau_f$, which characterizes the terminal point of the trajectory.
In the case of targeting an orbit, $\tau_f$ describes the time required to reach a fixed point on the orbit. 
When targeting the stable invariant manifold of a halo orbit, $\tau_f$ corresponds to the coast time needed to reach the insertion point.
After propagating the initial state (fixed) and initial costate (from the ACT method) according to the system dynamics, we determine the optimal terminal point for the given initial costate guess.
To do so, for each state along the propagated trajectory, the nearest point on the target orbit or manifold is identified by iterating over the time of flight.
This involves a nearest neighbor search based on the Euclidean distance of position and velocity\footnote{Since $\boldsymbol{r}$ and $\boldsymbol{v}$ are normalized using natural units we use the combined $\left\Vert \cdot \right\Vert_\infty$-norm of both. In other cases these should be treated separately.}, which is conducted using a k-dimensional (k-d) tree \cite{Bentley.1975}.
This k-d tree only has to be generated once during the problem setup and partitions the states of the target orbit or manifold in a tree structure.
It allows the nearest neighbor search for a given trajectory to be conducted more efficiently than conventional search algorithms, by moving down the tree structure recursively.
After calculating the constraint violations relative to the closest target point for each state along the trajectory, the minimum constraint violation $\delta_{c,min}$ and the corresponding shooting time $\tau_s$ and final coast time $\tau_f$ are identified.

If the constraint violation exceeds the chosen global search tolerance $\delta$, the algorithm switches to the next initial guess without performing optimization.
For the case $\delta_{c,min}<\delta$, the local optimization routine is initialized with $\overline{\boldsymbol{u}}$.
If the terminal constraint violation falls below the global search tolerance $\delta$, the corresponding $\tau_s$ and $\tau_f$ are determined, indicated by a "+" symbol next to the triangles in Fig. \ref{fig: global search}.
These approximate solutions $\overline{\boldsymbol{u}}$ can be refined to local optimality using SNOPT.
Since the preliminary screening algorithm already determines an ideal terminal point with corresponding shooting time, we use a fixed terminal boundary condition for the local optimization.
An alternative approach, shown as the bottom path in the figure, sets $\delta$ equal to the desired problem convergence tolerance. 
This generates feasible solutions without optimization by filtering initial guesses from the ACT method.
Due to the high efficiency of the preliminary screening algorithm, this strategy can generate feasible solutions at a high rate, despite only a small fraction of initializations passing the screening.
However, the drawback of this method is that the rate of solutions declines as problem complexity increases and the convergence tolerance becomes more stringent.
\subsection{Generative Learning Framework}
\begin{figure}[t!]
	\centering
	\includegraphics[width=0.9\textwidth]{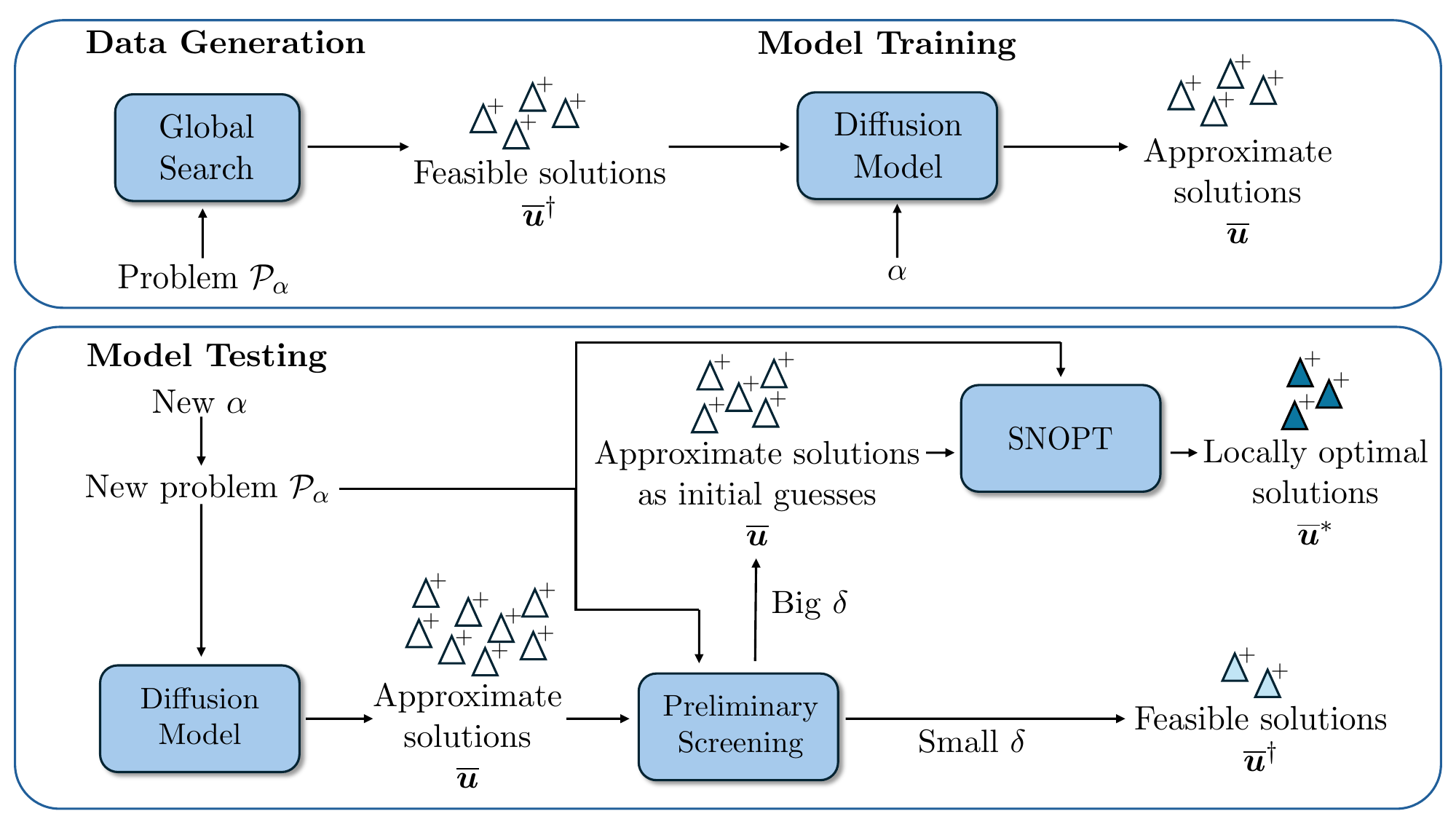}
	\caption{Visualization of the workflow used to train the diffusion model with fixed conditional parameters $\alpha$ and to test it on unseen $\alpha$-values.}
	\label{fig: diffusion model framework}
\end{figure}
The framework used in this work to accelerate the global search problem of low-thrust trajectory optimization with generative machine learning is based on the Amortized Global Search (AmorGS) framework introduced by Li et al. \cite{Li.872023}.
The process, illustrated in Fig. \ref{fig: diffusion model framework}, consists of three main steps.
First, a large dataset of feasible or locally optimal solutions is generated using the indirect method combined with the global search strategy. 
These solutions correspond to a finite set of problems $\{\mathcal{P}_{\alpha}\}$, where the conditional parameter $\alpha$ is sampled from a bounded set (e.g., $\alpha \in [\underline{\alpha},\overline{\alpha}]$).
Next, the generated decision vectors $\overline{\boldsymbol{u}}$ and their associated $\alpha$ are used to train the generative machine learning model, as shown at the top of Fig. \ref{fig: diffusion model framework}.
In this paper a diffusion model is used for this purpose. 
A successful training of the model results in it learning how the solution space distribution (visualized in Fig. \ref{fig:Diffusion Model}) varies with $\alpha$ and enables prediction of the structure for unseen values of the conditional variable.
In the final step, labeled \textit{Model Testing} in Fig. \ref{fig: diffusion model framework}, the diffusion model generates initial guesses for a problem $\mathcal{P}_{\alpha}$, often corresponding to a conditional variable $\alpha$ that wasn't encountered in the data generation and training phase. 
Ideally, these sampled indirect decision vectors $\overline{\boldsymbol{u}}$ serve as good approximations for unknown locally optimal solutions.
Work to incorporate constraint satisfaction with diffusion models is nascent and not straightforward (see Li et al. \cite{Li:2024.conf.dddas} for some efforts on aerospace and robotic related problems). 
Hence, the initial guesses generated by the diffusion model may not fully meet the desired feasibility tolerances.  
The initial guesses are therefore combined with the preliminary screening algorithm, which also calculates the shooting and final coast times $\tau_s$ and $\tau_f$ as described in Fig. \ref{fig: preliminary screening}. 
This has the additional benefit of the model not having to learn $\tau_s$ and $\tau_f$, therefore reducing the dimensionality of its input vector.
After screening, the solutions can either be refined to local optimality using SNOPT (upper path in Fig. \ref{fig: diffusion model framework}) or directly used as feasible solutions without further optimization by setting the global search tolerance to the desired feasibility tolerance (lower path in Fig. \ref{fig: diffusion model framework}).

The initial stages of generating data (5,000-10,000 CPU-hours), training the model (3-4 GPU-hours), and sampling from it (5-6 GPU-minutes) are time-consuming and would not be necessary if we directly solve the target problem. 
However, during preliminary mission design, the specific value of the conditional parameter is not fixed yet, requiring rapid generation of solutions across wide parameter ranges.
As a result, it may be worth investing in the upfront effort to enable quick exploration and adaptation during early mission design phases or even in real-time during a mission to respond to unforeseen events.

\subsection{Diffusion Model}
Generative models have gained significant popularity in recent years, due to their ability to provide samples from complex distributions that characterize datasets ranging from images, audio, video, and language. 
The models must therefore learn the hidden structure of a distribution that represents the training dataset. 
In practice, all generative models follow two steps: first a sample is drawn from a distribution that is easy to sample from (e.g., a standard multivariate Gaussian) and second, the sample is mapped to a corresponding sample from the distribution of interest. 
Diffusion probabilistic models, or simply diffusion models (DM), which we use in this paper, sample from a standard Gaussian distribution and then in the second step learn a time reversed diffusion process.
Reverse-time diffusion equations were first studied by Anderson \cite{Anderson:1982.spa.12.3} and then Haussmann and Pardoux \cite{Haussmann:1986.ap.14.4}. 
These ideas, unknown to the original machine learning contributors, were introduced to machine learning for image generation by Sohl-Dickstein et al. \cite{SohlDickstein.3122015}.
Diffusion model capabilities in both conditional and unconditional image, audio, and video generation have since been greatly improved by Ho et al. \cite{Ho.6192020}, and Song and Ermon \cite{Song.7122019}. 
Following their success in image generation, diffusion models have been applied to areas like reinforcement learning \cite{Wang.8122022,Ding.252024,Ding.9292023}, motion generation \cite{Tevet.9292022}, and robotics \cite{Chi.372023}.

Diffusion models offer several advantages over other common generative learning models, making them particularly attractive for trajectory optimization problems. 
Unlike other flow-based models, diffusion models are less constrained in their architecture, providing greater flexibility in model design and enhancing their ability to capture complex data distributions effectively \cite{Zhang.2021}. 
Recent studies have shown that diffusion models outperform generative adversarial networks (GANs) in image synthesis due to more stable training and higher diversity in generated data \cite{Dhariwal.5112021}.
Additionally, compared to conditional variational autoencoders (CVAEs), diffusion models have demonstrated significantly higher sample quality for trajectory optimization problems \cite{Li.2222024}.

\begin{figure}[t!]
	\centering
    \includegraphics{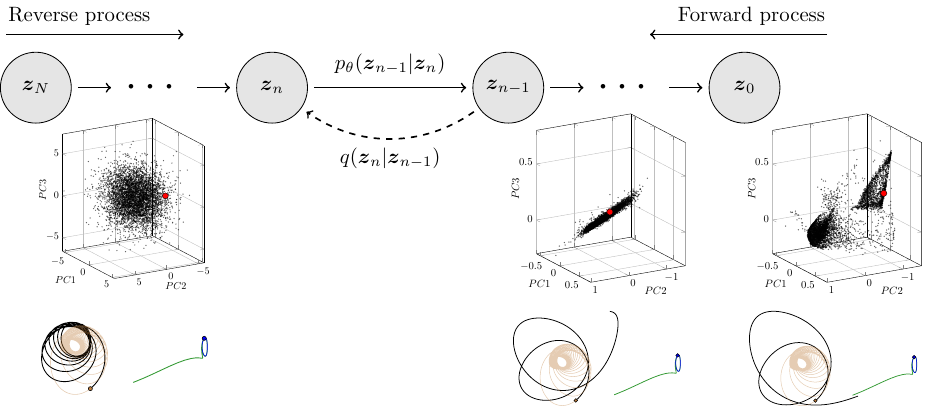}
    \caption{Illustration of the application of diffusion models, in the context of indirect low-thrust trajectory optimization. The trajectories resulting from a selected costate vector (\textcolor{red}{red}) from the distribution are shown at three diffusion steps.} 
	\label{fig:Diffusion Model}
\end{figure}

Generation of samples using a diffusion model involves a time-reversed diffusion process, but to train the model to learn this reverse process, a forward diffusion process is also required. 
Figure \ref{fig:Diffusion Model} illustrates the idea of the forward and reverse process. 
In particular, given samples $\boldsymbol{z}_0$ of the training dataset, these are evolved under a user-prescribed forward diffusion, approximated with a fixed step integration scheme, to a final sample $\boldsymbol{z}_N$ after $N$ integration steps. 
The forward diffusion is chosen to yield final samples from a "nice" distribution (i.e., one that will be easy to sample from during the generation process, such as a Gaussian distribution). 
The reverse process then learns the time-reversal of the forward diffusion equation, so that samples can be drawn from the "nice" distribution and mapped to the desired and likely complex distribution. 
\label{sec: Diffusion models}
In Fig. \ref{fig:Diffusion Model}, this concept is illustrated for the global costate structure of a spacecraft trajectory optimization problem that will be described in the GTO Halo section of this paper. 
The trajectory corresponding to a selected data point from the costate distribution is visualized.
In particular, the physical realization (i.e., simulation using \texttt{pydylan}) is illustrated at the bottom of the figure. As the sample is progressed through the reverse process, the feasibility of the trajectory is improved, until at $\boldsymbol{z}_0$ the sample has very little constraint error, such that an NLP solver can quickly converge to a locally optimal solution. 
Additional details of these figures will be provided in the results section.

\subsubsection{Forward Process}
Perhaps the simplest choice for a forward diffusion process is given by the discretization of an Ornstein-Uhlenbeck process, 
\begin{align}
\boldsymbol{dz}_t = -\gamma \boldsymbol{z}_t dt + \boldsymbol{\sigma}_t \boldsymbol{dw}_t, \quad \boldsymbol{z}_0 \sim \mathbb{P}_\alpha, \quad t > 0,
\end{align}
where $\mathbb{P}_\alpha$ is the distribution for the problem specified by the conditional parameter $\alpha$, $\boldsymbol{dw}_t$ is standard Brownian motion, $\boldsymbol{\sigma}_t \succ 0, \forall t > 0$, the dispersion coefficient (a positive definite matrix) and $\gamma > 0$ the drift coefficient. 
Because $-\gamma$ provides dissipation and the Brownian motion is non-degenerate, the limiting distribution given by $\boldsymbol{z}_t$ is a standard Gaussian distribution. 
This stochastic differential equation is also a continuous-time Markov process. 

As stated earlier, diffusion models approximate the continuous-time forward process with a fixed-step numerical integration, and the number of integration steps, $N$, is chosen along with a schedule for the dispersion coefficient and choice of drift coefficient such that $\boldsymbol{z}_N \sim \mathcal{N}(0, \boldsymbol{I})$ (i.e., from a standard Gaussian, where $\boldsymbol{I}$ is the identity matrix). 
This is the approach we follow, and in particular we chose a scheduling of the dispersion influenced by the work of Ho et al. \cite{Ho.6192020} on denoising diffusion probabilistic models.
The choice of $\boldsymbol{\sigma}_t$ now becomes a sequence $(\beta_n)$, and to provide a closed form relation for the diffusion, the drift is chosen dependent on this schedule such that we have, 
\begin{align} 
\label{eq: forward diffusion process step}
q(\boldsymbol{z}_n | \boldsymbol{z}_{n-1}) = \mathcal{N}(\boldsymbol{z}_n; \sqrt{1 - \beta_n}\boldsymbol{z}_{n-1}, \beta_n \boldsymbol{I}),
\qquad
q(\boldsymbol{z}_{1:N} | \boldsymbol{z}_{0}) = \prod_{n=1}^N q(\boldsymbol{z}_n | \boldsymbol{z}_{n-1}).
\end{align}
The sequence $(\beta_n \in (0, 1))$ is chosen according to a cosine-based schedule, which controls the rate at which information is diluted by the driving Brownian motion.
For training purposes (i.e., backpropagation), it is necessary to use reparameterization\footnote{often referred to as the reparameterization trick in the machine learning literature, which is simply the invariance property of a Gaussian random variable under affine transformation}, which allows to directly sample at any arbitrary timestep $n \in \{1,2, \hdots ,N\}$,
\begin{equation}
\label{eq: forward process reparameterized}
\boldsymbol{z}_n =\sqrt{\overline{\alpha}_n} \boldsymbol{z}_0 + \sqrt{1 - \overline{\alpha}_n} \boldsymbol{\epsilon},
\end{equation}
where $\boldsymbol{\epsilon} \sim \mathcal{N}(\mathbf{0},\boldsymbol{I})$, $\alpha_n = 1 - \beta_n$ and $\overline{\alpha}_n = \prod_{i=1}^n \alpha_i$.
When the dilution of information is sufficiently small during the structured integration steps, then a stable recovery during the reverse process can be accomplished.

\subsubsection{Reverse Process}

In diffusion models, a neural network is used to approximate the true time-reversal transition probability for a given step, 
\begin{align}
\label{equation: NN approximation of time-reversal transition density}
p_\theta(\boldsymbol{z}_{n-1} | \boldsymbol{z}_n) \approx q(\boldsymbol{z}_{n-1} | \boldsymbol{z}_n).
\end{align}
A neural network is a flexible mathematical model built from layers of simple processing units (“neurons”).
The parameters of the network, which we denote by $\theta$ in Eq. \eqref{equation: NN approximation of time-reversal transition density}, are tuned during training to specify their connection weights and learn to transform inputs into accurate outputs.
Assuming that the diffusion for any given step is sufficiently small (e.g., by choosing a sufficiently high number of integration steps and an appropriate dispersion coefficient schedule), it is appropriate to model the transition probability as a Gaussian distribution with conditional mean $\boldsymbol{\mu}_\theta$,
\begin{align} 
\label{eq: reverse diffusion process}
p_{\theta}(\boldsymbol{z}_{n-1} | \boldsymbol{z}_n) = \mathcal{N}(\boldsymbol{z}_{n-1}; \boldsymbol{\mu}_{\theta}(\boldsymbol{z}_n, n), \tilde{\beta}_n\boldsymbol{I}),
\end{align}
and $\tilde{\beta}_n = (1 - \overline{\alpha}_{n-1})/(1 - \overline{\alpha}_n)\cdot \beta_n $.
The neural network takes $\boldsymbol{z}_n$ and $n$ as input and outputs $\boldsymbol{\mu}_{\theta}$.
The goal of training a diffusion model then becomes the standard problem of maximizing the likelihood of the data for the given probability distribution $p_\theta(\boldsymbol{z}_0)$, which due to the Markovian property becomes a product of conditional Gaussian distributions. 
As is common in generative machine learning, a variational bound is instead extremized, since it has better numerical tractability.
This variational bound is composed of two terms, a Kullback-Leibler (KL) divergence and a reconstruction loss. 
In the Gaussian setting, the KL divergence is computable with knowledge of the determinants of the covariances, which is most easily done when the covariances are only non-zero on the diagonal. 
If sample sizes are large enough, the KL term is less significant. 
Following this assumption, similar to Ho et al. \cite{Ho.6192020} and Sohl-Dickstein et al. \cite{SohlDickstein.3122015}, we define the loss function to be extremized, and hence guide the learning of the DM, as just the reconstruction loss, 
\begin{equation} 
\label{eq: simplified loss}
L(\theta) \equiv \mathbb{E}_{n \sim [1, N], \boldsymbol{z}_0, \boldsymbol{\epsilon}_n} \left[ \left\| \boldsymbol{\epsilon}_n - \boldsymbol{\epsilon}_{\theta}(\boldsymbol{z}_n, n) \right\|^2 \right],
\end{equation}
where the expectation is over all possible diffusion steps in the reverse process, $\boldsymbol{\epsilon}_n$ is the known diffusion from the forward process, and $\boldsymbol{\epsilon}_{\theta}(\boldsymbol{z}_n, n)$ is the learned reverse diffusion noise. 
The entire step-by-step process is detailed in the work of Sohl-Dickstein et al. \cite{SohlDickstein.3122015}. 
A neural network based on a U-Net architecture, originally introduced by Ronneberger et al. \cite{Ronneberger.5182015} is trained on this loss using a stochastic gradient descent (SGD) algorithm.
In SGD, the model’s parameters are updated iteratively using an approximation to the true expected gradient by computing an averaged one using a small, randomly selected mini-batch of gradients associated with the training data. 
In each iteration, the mini-batch is fed forward through the network to produce predictions, the loss between these predictions and the ground truth is calculated, and the resulting error gradients are back-propagated to adjust the network weights.

\subsubsection{Classifier Free Guidance}
Classifier free guidance is a simple approach introduced by Ho et al. \cite{Ho.2022}, which allows to generate diffusion model outputs based on specific conditions. 
The key idea is to train a single neural network to produce both conditional and unconditional samples.
By discarding the conditional variable with a fixed probability (i.e., setting it to the null element $\alpha = \emptyset$),
the unconditional model can be parameterized using the same neural network. 
During sampling a linear combination of the conditional and unconditional score estimates is used:
\begin{equation}
\label{eq: classifier free guidance}
\overline{\boldsymbol{\epsilon}}_{\theta}(\boldsymbol{z}_n, n, \alpha) = (w + 1)\boldsymbol{\epsilon}_{\theta}(\boldsymbol{z}_n, n, \alpha) - w \boldsymbol{\epsilon}_{\theta}(\boldsymbol{z}_n, n, \alpha = \emptyset),    
\end{equation}
with the guidance strength $w>0$ controlling the trade-off between sample variety and individual sample fidelity. 
The new score estimate $\overline{\boldsymbol{\epsilon}}_{\theta}$, is then used to predict $\boldsymbol{z}_{n-1}$:
\begin{equation} 
\label{eq: sampling}
\boldsymbol{z}_{n-1} = \frac{1}{\sqrt{\alpha_n}} \left( \boldsymbol{z}_n - \frac{1 - \alpha_n}{\sqrt{1 - \alpha_n}} \overline{\boldsymbol{\epsilon}}_{\theta} \right) + \sqrt{\tilde{\beta}_ n}\boldsymbol{y}(n),
\end{equation}
where $\boldsymbol{y}(n) \sim \mathcal{N}(\boldsymbol{0}, \boldsymbol{I})$ if $n > 1$ and $\boldsymbol{y}(1) = 0$;
that is, this sampling loop runs from $n = N$ to $n = 1$, with no noise being added in the last step to generate the final sample $\boldsymbol{z}_0$.
Since the model does not impose any bounds on the generated $\boldsymbol{z}_0$, control values outside their lower or upper bounds are clipped.

\section{Results}
In this section, the strong performance of the diffusion model framework is presented for two example missions within the CR3BP system. 
The first mission involves a planar DRO transfer in the Jupiter-Europa system.
This problem was chosen because it has a smooth front of Pareto-optimal solutions found by Russell \cite{Russell.2007}, which are likely to be good numerical approximations of the true Pareto front.
To demonstrate the framework on a more complex scenario, we next examine a three-dimensional transfer in the Earth–Moon system. 
This second case is a modified version of the transfer solved by Li et al.\cite{Li.872023}, for which no clear Pareto front has been found yet.
\subsection{Europa DRO Transfer}
\label{sec: Europa DRO}
\subsubsection{Problem Description}
\label{section: problem description}
\begin{figure}[t!]
    \centering
    \begin{minipage}{0.5\textwidth}
        \centering
        \includegraphics[scale=0.7]{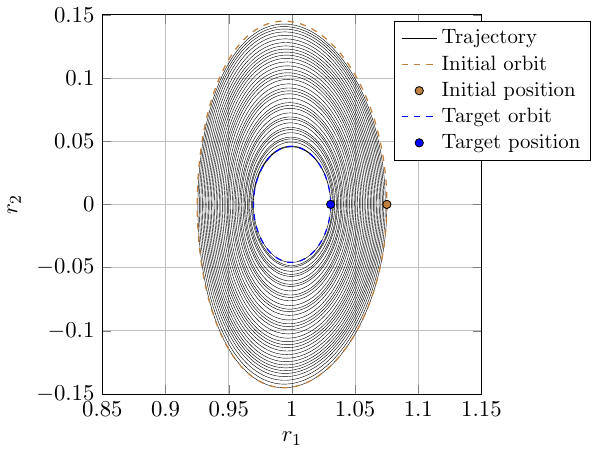}
        \caption{Trajectory with $T_{max} = 0.9968\,\unit{N} (\alpha=0.2)$ for the Europa DRO transfer in the CR3BP frame.}
        \label{fig:example trajectory}
    \end{minipage}
    \hfill
    \begin{minipage}{0.47\textwidth}
        \centering
        \includegraphics[scale=0.75]{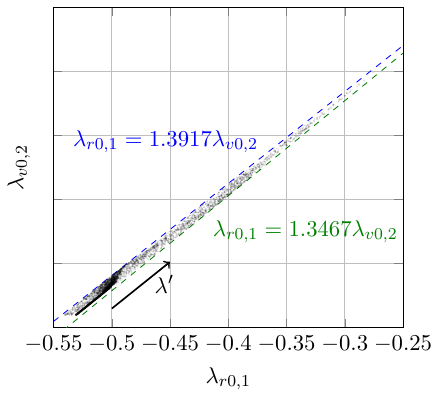}
        \caption{Linear relationship between the costates $\lambda_{r0,1}$ and $\lambda_{v0,2}$ for $\alpha=1.0$ $(T_{max}=4.984\,\mathrm{N})$.}
        \label{fig:lamr1v2}
    \end{minipage}
\end{figure}

The first problem is a planar low-thrust transfer in the Jupiter-Europa CR3BP system. 
It derives from the Jupiter Icy Moons Orbiter (JIMO) mission concept, which was canceled in 2005.
The spacecraft starts in an initial Distant Retrograde Orbit (DRO) at Europa and transfers to a lower DRO.
The family of DROs around Europa evolves from planar, near circular orbits close to Europa to bigger, almost elliptical orbits with Europa at the center.
Hence, this is a planar transfer, although the spacecraft is not explicitly restricted to stay in this plane by $\texttt{pydylan}$. 
The initial and final orbits, along with an example trajectory, are shown in Fig. \ref{fig:example trajectory}.
The spacecraft starts at a fixed position on the initial orbit, located at the $r_1$-axis crossing farther from Jupiter. 
The phase at which the spacecraft transfers to the target orbit is also free and is represented by a final coasting time $\tau_f \in [0, \mathcal{T}_f)$, where $\mathcal{T}_f$ is the orbital period of the target orbit. 
A final coast time of $\tau_f = 0$ corresponds to an orbit insertion at the $r_1$-axis crossing of the target orbit, marked as the terminal state in Fig. \ref{fig:example trajectory}.
\begin{table}[t!]
\centering
\caption{Problem parameters for Europa DRO transfer.} 
\setlength{\tabcolsep}{8pt} 
\begin{tabular}{lclc} 
\toprule
\multicolumn{2}{l}{\textbf{Trajectory parameters}} & \multicolumn{2}{c}{} \\
\midrule
\multicolumn{2}{l}{Initial state $[\boldsymbol{r}_0^T,\boldsymbol{v}_0^T]$ [NU]} & \multicolumn{2}{c}{$[1.0752, 0.0, 0.0, 0.0, -0.1499, 0.0]$}\\
\multicolumn{2}{l}{Terminal state $[\boldsymbol{r}_f^T,\boldsymbol{v}_f^T]$ [NU]} & \multicolumn{2}{c}{$[1.0306, 0.0, 0.0, 0.0, -0.0727, 0.0]$}\\
\multicolumn{2}{l}{Orbital period target DRO $\mathcal{T}_f$ [NU]} & \multicolumn{2}{c}{$4.1055$}\\
\midrule
\multicolumn{2}{l}{\textbf{Spacecraft parameters}} & \multicolumn{2}{l}{\textbf{Natural units (Jupiter-Europa)}} \\
\midrule
Initial mass $m_0$ [kg] & 25,000 & Distance unit [km] & 670,900 \\
Fuel mass [kg] & 15,000 & Time unit [s] & 48,822.76 \\
Dry mass [kg] & 10,000 & Mass unit [kg] & $1.898 \times 10^{27}$ \\
Specific impulse $I_{sp}$ [s] & 7,365 & & \\
Thrust range $T_{max}$ [N] & $[0.4984, 4.984]$ & & \\
\bottomrule
\end{tabular}
\label{tab:problem parameters Europa DRO}
\end{table}
All relevant parameters for this transfer are presented in Table \ref{tab:problem parameters Europa DRO}.
The spacecraft parameters correspond to those used by Russell \cite{Russell.2007}, with the only difference being that the maximum engine thrust is varied as a problem parameter. 
All variables in this section are presented in natural units of the Jupiter-Europa system unless stated otherwise. 
The values used to normalize the units of distance, time and mass are given in Table \ref{tab:problem parameters Europa DRO}.

\subsubsection{Data Generation}
\label{section: data generation}
The goal for this problem is to train a diffusion model based on the conditional parameter $\alpha$, which describes the spacecraft’s maximum thrust level. 
The relationship between $T_{max}$ and $\alpha$ is defined by: {$\alpha = T_{max}/(4.984\,\mathrm{N})$},
with $\alpha\in[0.1,1.0]$.
To train the model, a dataset is created, consisting of solutions for ten fixed thrust levels, which equidistantly cover the defined $\alpha$-range.
The dataset includes 270,000 solutions, with 27,000 solutions per thrust level.
Directly combining the preliminary screening algorithm with a tolerance $\delta=1\times10^{-4}$ [NU] proves to be a highly efficient approach for quickly generating feasible solutions to this problem. 
Consequently, the strategy corresponding to the bottom path in Fig. \ref{fig: global search} is used.
Although this tolerance is coarse by the standards of mission design, it allows us to capture the global solution distribution efficiently.
Russell \cite{Russell.2007} demonstrated that there exists a well-defined front of Pareto optimal solutions in the $\Delta v$-time-of-flight plane for this problem with $\alpha=1.0$.
Solutions are labeled Pareto optimal if no other solutions have both a lower change in velocity ($\Delta v$) and a shorter time of flight. 
These solutions are specifically targeted in the data generation process, as they represent the most interesting trajectories from a mission design perspective.
The global search parameters used for data generation are displayed in Table. \ref{tab: GS parameters Europa DRO}.
The ranges for $\varphi$, $\dot{\varphi}_0$, $\beta_0$, $\dot{\beta}_0$, $S_0$ and $\dot{S}_0$ were chosen based on the Pareto optimal solutions found by Russell \cite{Russell.2007} and further refined through test runs.
\begin{table}[b!]
\centering
\caption{Global search parameters for the Europa DRO transfer.}
\setlength{\tabcolsep}{8pt} 
\begin{tabular}{p{5cm}c|p{2cm}c} 
\toprule
\textbf{Parameter} & \textbf{Value/Range} & \textbf{Parameter} & \textbf{Value/Range} \\
\toprule
Global search tolerance $\delta$ [NU] & $\num{1e-4}$ & $\varphi_0$ [rad] & $[\pi-0.012,\pi+0.01]$ \\
Feasibility tolerance [NU] & $\num{1e-4}$ & $\dot{\varphi}_0$ [rad/TU] & $[-0.02,0.025]$ \\
SNOPT runtime limit [s] & $30$ & $S_0$ [NU] & $[0.0,0.2]$ \\
SNOPT mode & optimal & $\dot{S}_0$ [NU] & $[-0.0022,0.004]$ \\
\bottomrule
\end{tabular}
\label{tab: GS parameters Europa DRO}
\end{table}

\begin{figure}[t!]
    \centering
    \begin{subfigure}[b]{0.34\textwidth}
        \centering
        \includegraphics[scale=0.65]{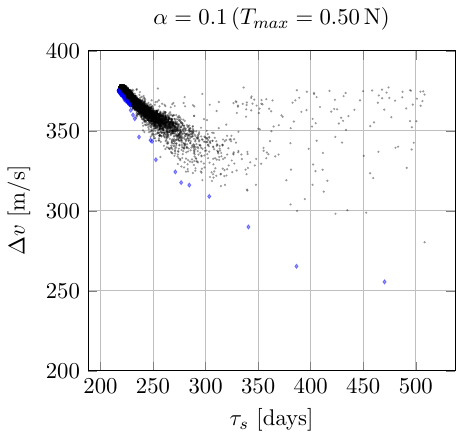}
    \end{subfigure}\hfill
        \begin{subfigure}[b]{0.33\textwidth}
        \centering
        \includegraphics[scale=0.65]{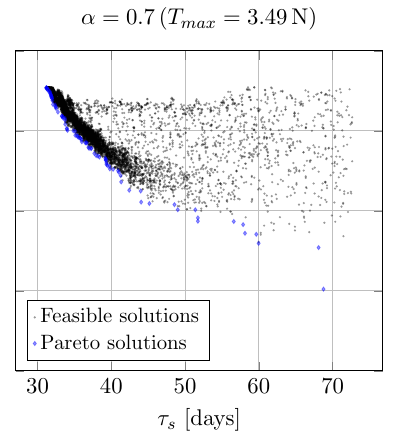}
    \end{subfigure}\hfill
    \begin{subfigure}[b]{0.33\textwidth}
        \centering
        \includegraphics[scale=0.65]{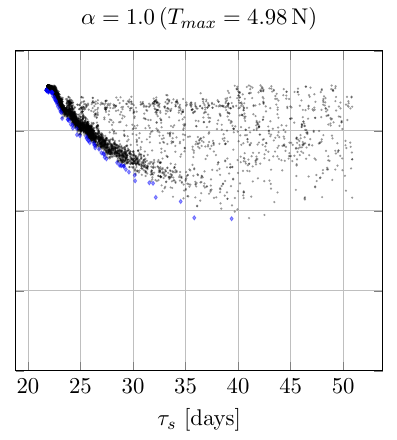}
    \end{subfigure}
    \caption{A subset of 10,000 solutions from the training dataset for each of three thrust levels $\alpha$, displayed in the $\Delta v$-$\tau_s$ frame.}
    \label{fig: dv-TOF plots}
\end{figure}
The generated dataset primarily consists of solutions near the respective Pareto front for each $\alpha$, as demonstrated in Fig. \ref{fig: dv-TOF plots}. 
A subset of solutions for three different $\alpha$ values is displayed in the plane of the competing objectives, $\Delta v$ and shooting time $\tau_s$. 
The displayed Pareto front consists of the Pareto-approximate solutions from this subset for each thrust level.  For $\alpha = 1.0$, the Pareto front found by Russell \cite{Russell.2007} is reproduced. 
The lower thrust levels reveal a similar structure, with comparable $\Delta v$ and longer times of flight.
For $\alpha=0.1$ the solutions tend to cluster more towards the upper part of the Pareto front, as generating solutions with longer flight times appears to be more challenging at very low thrust levels. 
However, these solutions are still valuable for mission design due to their comparatively shorter times of flight.

\subsubsection{Solution Structure}
\label{section: solution structure}
Analyzing the solutions in the generated dataset reveals clusters in the costate space that vary depending on the thrust magnitude $T_{max}$. 
These clusters likely correspond to basins of locally optimal trajectories in the solution space. 

A linear relationship is found between $\lambda_{r0,1}$ and $\lambda_{v0,2}$, which is shown in Fig. \ref{fig:lamr1v2} and can be explained by evaluating the equations used for the ACT method. 
The thrust directions at the initial time in the CR3BP frame are determined using the transformation presented in Eq. \ref{eq: thrust transformation}, with $\beta_0=0$ and $\dot{\beta}_0=0$.
Additionally, based on the ranges from Table \ref{tab: GS parameters Europa DRO}, it is $S_0>0$ and we assume $\alpha\approx\pi$.
The transformation matrices $R_0$ and $\dot{R}_0$ are sparse due to the initial conditions $r_{0,2}=0$ and $v_{0,1}=0$, resulting in $\hat{\boldsymbol{u}}_0\approx(0,1,0)$ and $\dot{\hat{\boldsymbol{u}}}_0\approx(0.628 + \dot{\alpha}_0,0,0)$.
From Eq. \eqref{Equation: Control Law}, Eq. \eqref{Equation: Costate equation of motion}, Eq. \eqref{Equation: ACT1} and Eq. \eqref{Equation: ACT2}, it follows that:
\begin{equation}
    \label{eq: linear costate derivation}
    \boldsymbol{\lambda}_{r0}=(\dot{S}_0 - \lambda_{m0} \frac{\dot{m}_0}{c} + \lambda_{v0} \frac{T_{max}}{m_0c}) \hat{\boldsymbol{u}}_0 + \lambda_{v0} \dot{\hat{\boldsymbol{u}}}_0 - \boldsymbol{H}^T\boldsymbol{\lambda}_{v0},
\end{equation}
where the index indicates that variables that are not constant along the trajectory are evaluated at $t=0$.
From $\boldsymbol{\lambda}_{v0} = -\lambda_{v0} \hat{\boldsymbol{u}}_0$ follows that $\boldsymbol{\lambda}_{v0,2} >> \boldsymbol{\lambda}_{v0,1}$ and therefore $\lambda_{v0}\approx |\lambda_{v0,2}|$.
Simplifying Eq. \eqref{eq: linear costate derivation} yields the explanation for the linear relationship in Fig. \ref{fig:lamr1v2}:
\begin{align}
    \label{eq: linear costate derivation simple}
    \lambda_{r0,1} &\approx -(0.6283 - \dot{\alpha}_0) \lambda_{v0,2} + 2\lambda_{v0,2} = (1.3692\pm0.0225)\lambda_{v0,2},
\end{align}
The expansion perpendicular to the slope is caused by the specific value of $\dot{\alpha_0}$ from the given range in Table \ref{tab: GS parameters Europa DRO}.
The two straight lines with a slope, corresponding to the boundary of the $\dot{\alpha_0}$ interval, are shown as blue dashed lines in Fig. \ref{fig:lamr1v2}.
All points in that plot fall between these two lines, confirming the accuracy of the mathematical explanation for this relationship.
For visualization purposes, a new transformed costate variable, $\lambda'$, is introduced, based on this linear equation:
\begin{equation}
    \label{eq: lambda_trans}
    \lambda' = \arctan(\cos(1.3692))\lambda_{r0,1}+\arctan(\sin(1.3692))\lambda_{v0,2}= 0.5895\lambda_{r0,1}+0.8075\lambda_{v0,2}.
\end{equation}
The new variable consolidates the linear dependency into a single dimension by capturing the maximum variance in those costates. 
The direction of the $\lambda'$ axis is indicated by the black arrow in the bottom right plot of Fig. \ref{fig:lamr1v2}.
The transformed costate variable permits a visualization of the two position and two velocity costates in a three-dimensional plot.
\begin{figure}[t!]
    \centering
    \begin{subfigure}[b]{0.33\textwidth}
        \centering
        \includegraphics[scale=0.63]{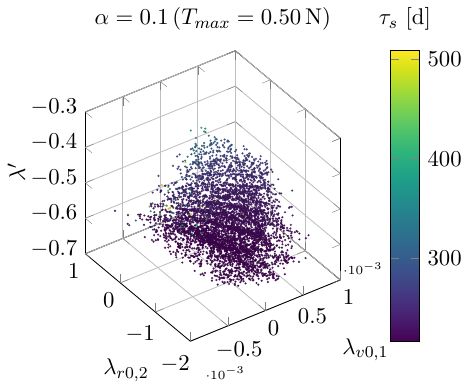}
    \end{subfigure}\hfill
        \begin{subfigure}[b]{0.33\textwidth}
        \centering
        \includegraphics[scale=0.63]{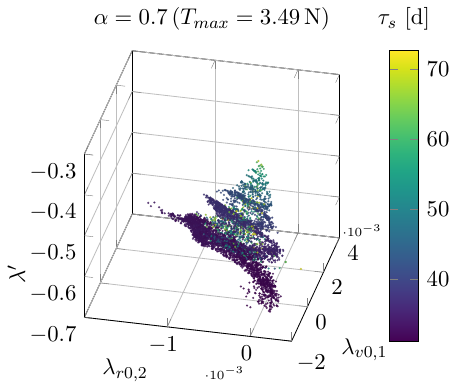}
    \end{subfigure}\hfill
    \begin{subfigure}[b]{0.33\textwidth}
        \centering
        \includegraphics[scale=0.63]{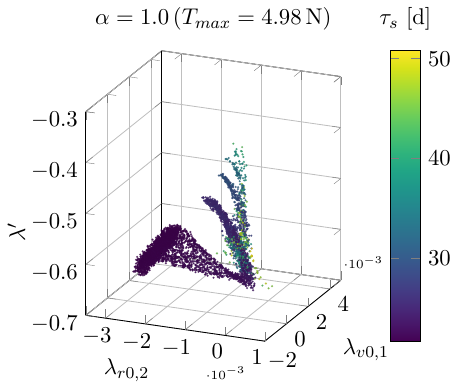}
    \end{subfigure}
    \caption{Hypersurface structure in the costate space for three different thrust levels $\alpha$. Each plots displays a subset of 5,000 solutions from the training data.}
    \label{fig: lamr2v1_trans}
\end{figure}
The resulting curved hypersurface structure can be observed in Fig. \ref{fig: lamr2v1_trans} for three different thrust levels. 
The shape and location of these curved surfaces vary depending on the thrust level. 
A clear structure is evident for $\alpha = 1.0$, but as the thrust level decreases, the structure becomes more diffuse, with the hypersurfaces barely recognizable for $\alpha = 0.1$. 
Additionally, the shooting time $\tau_s$ is visualized through a color plot, revealing a trend where hypersurfaces with higher $\lambda'$ values correspond to longer shooting times.
However, many solutions with longer shooting times do not fit into the hypersurface structure.
\subsubsection{Diffusion Model Training}
\label{section: diffusion model training}
The neural network used to model one step in the reverse process of the denoising diffusion probabilistic model from Ho et al. \cite{Ho.6192020} is based on a U-Net architecture, originally introduced by Ronneberger et al. \cite{Ronneberger.5182015}. 
A U-Net consists of an encoder and a decoder. 
The encoder compresses the input data into a lower-dimensional hidden representation, resulting in a bottleneck layer. 
After the bottleneck, the decoder restores this representation to the original input dimension, with residual connections from the encoder to the decoder enhancing gradient flow.
The diffusion model parameters
that were used for training and sampling are presented in Table \ref{tab:layer_sizes}, with no claim for these parameters to be optimal for the given problem, since that would require an extensive hyperparameter study.
For reproducibility and completeness we provide extra details on the network architecture and a description of all hyperparameters in the Appendix.
In the following, a brief explanation of some of the trade-offs and comparisons made to justify the final parameter selection is given.
\begin{table}[b!]
\centering
\caption{Hyperparameters of the diffusion model.}
\setlength{\tabcolsep}{8pt} 
\begin{tabular}{p{5.5cm}c|p{4cm}c} 
\toprule
\textbf{Parameter} & \textbf{Value} & \textbf{Parameter} & \textbf{Value} \\
\toprule
Sequence length $L$ & $6$ & Channel number $C$ & $1$ \\
Batch size $B$ & 1,024 & Encoder input dimension & $64$ \\
Encoder layer dimensions& $[256,256,512]$ & Timesteps $N$& 1,000\\
Embedded class layer dimension & $[128,256]$ & Training epochs& $200$\\
Time embedding dimension& $256$& Guidance strength $w$& $1.0$\\
Unconditional training probability & $0.1$ & &\\
\bottomrule
\end{tabular}
\label{tab:layer_sizes}
\end{table}

The batch size represents the subset of training data used during each step of the training process.
It controls how much GPU computational capacity and memory are used, directly impacting the training time. 
Increasing the batch size reduces training time until the GPU’s memory limit is reached.
However, a larger batch size also reduces variance in the generated samples, leading to lower sample diversity. 
The size of the U-Net influences the network’s capacity to capture and represent the complexity within the data structure. 
It is mainly determined through the number and dimensions of the encoder layers and embedded class layers, as well as the dimension of the time embedding.
Increasing the number of layers or their dimensions also leads to longer training and sampling times. 

\begin{figure}[t!]
    \centering
    \begin{minipage}[b]{0.35\textwidth}
        \centering
        \includegraphics[scale=0.61]{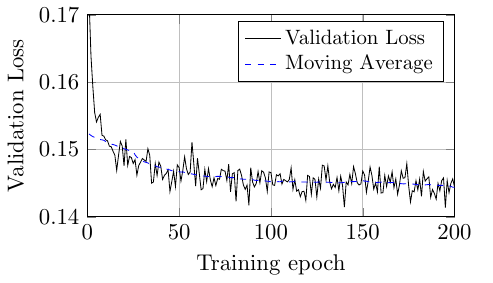}
        \caption{Evolution of the validation loss and its 50-epoch moving average during training.}
        \label{fig: validation_loss}
    \end{minipage}
    \begin{minipage}[b]{0.62\textwidth}
        \centering
        \begin{minipage}[b]{0.54\textwidth}
            \centering
            \includegraphics[scale=0.65]{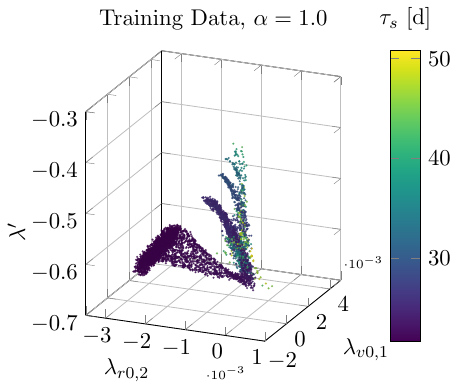}
        \end{minipage}\hfill
        \begin{minipage}[b]{0.46\textwidth}
            \centering
            \includegraphics[scale=0.65]{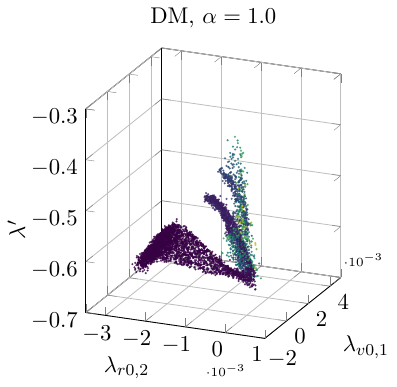}
        \end{minipage}
        \caption{Hypersurface structure in the costate space of the training data in comparison to diffusion model samples for \,$\alpha=1.0\,(T_{max}=4.98\,\mathrm{N})$.}
        \label{fig: lamr2v1_trans_training}
    \end{minipage}\hfill    
\end{figure}

The diffusion model is implemented and trained in PyTorch \cite{pytorch}.
Before training begins, the data is randomly split into a training set and a validation set in a 9:1 ratio.
The training set is the portion on which the model actually learns and updates its weights, while the validation set checks its performance on unseen data.
Training runs for a fixed number of epochs, where each epoch involves iterating through the entire training dataset.
The number of epochs is selected based on the validation loss, as shown in Fig. \ref{fig: validation_loss}. 
The moving average does not significantly decrease after 150 epochs, indicating that additional training beyond 200 epochs is unnecessary.
Choosing the number of diffusion timesteps, $N$, involves a trade-off between sampling quality and sampling time. 
The chosen value of $N$ = 1,000 results in a sampling time of $330.0$ s for 50,000 samples on a NVIDIA A100-SXM4-40GB GPU.
The diversity of the generated samples is controlled by the guidance strength $w$. 
If this value is set too high, the model primarily targets the largest clusters in the solution space and fails to generate initializations in the less densely populated regions. 
Conversely, setting $w$ too low reduces the model’s ability to incorporate classifier information during sample generation. 
\begin{figure}[t!]
    \centering
    \begin{subfigure}[b]{0.33\textwidth}
        \raggedright
        \includegraphics[scale=0.66]{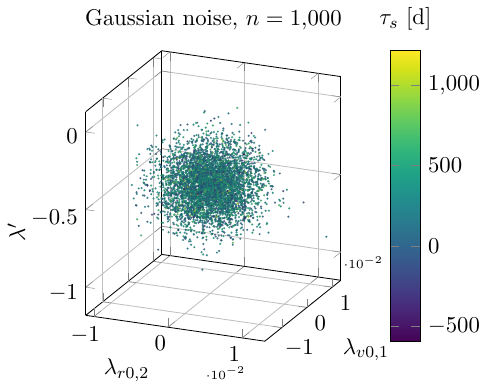}
    \end{subfigure}\hfill
    \begin{subfigure}[b]{0.33\textwidth}
        \raggedright
        \includegraphics[scale=0.66]{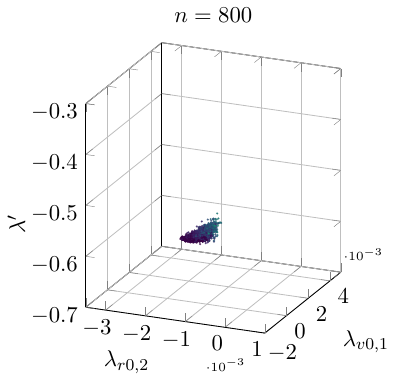}
    \end{subfigure}
    \begin{subfigure}[b]{0.33\textwidth}
        \raggedright
        \includegraphics[scale=0.66]{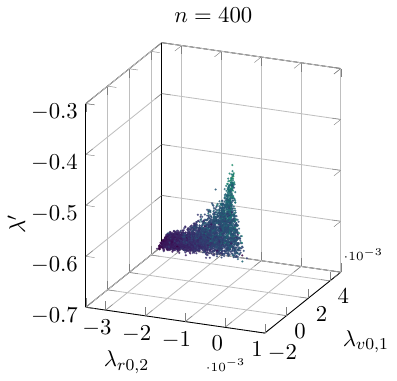}
    \end{subfigure}\hfill
    \begin{subfigure}[b]{0.33\textwidth}
        \raggedright
        \includegraphics[scale=0.66]{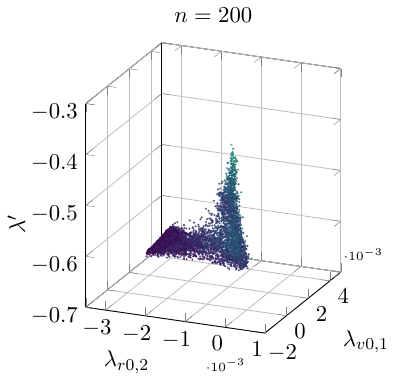}
    \end{subfigure}
    \begin{subfigure}[b]{0.33\textwidth}
        \raggedright
        \includegraphics[scale=0.66]{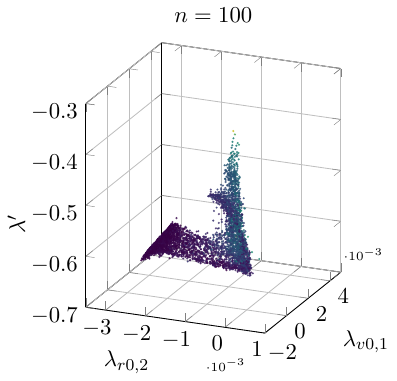}
    \end{subfigure}\hfill
    \begin{subfigure}[b]{0.33\textwidth}
        \raggedright
        \includegraphics[scale=0.66]{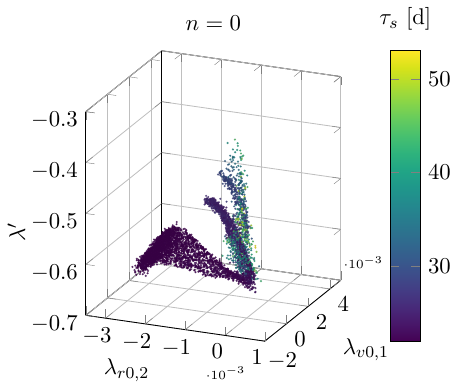}
    \end{subfigure}
    \caption{Structure in the costate space after different diffusion timesteps for thrust level $\alpha = 1.0\,(T_{max}=4.98\,\mathrm{N})$. Each plot displays a subset of 5,000 solutions. Note the different axis and colorbar scaling in the first plot, which shows the Gaussian noise at $n = 1$,000.}
    \label{fig: lamr1v2_DM_TS}
\end{figure}
The hyperparameters of the model were chosen based on a visual evaluation of the model's ability to reproduce the structure in the costate space for $\alpha=1.0$. 
While this is no quantifiable comparison, it is the best option to ensure that the costate structure is fully captured by our model.
A comparison of how this structure is captured by models with varying parameters is presented in the appendix. 
The reference model is shown in the first panel and each remaining panel varies in exactly one parameter, highlighting how that parameter impacts the distribution learned by the model.
Since $\alpha=1.0$ is part of the training dataset, a well-chosen set of parameters should enable the model to accurately generate a similar structure, starting from Gaussian noise.
This process is visualized for the final choice of parameters in Fig. \ref{fig: lamr1v2_DM_TS}.
The top left plot displays the initial data points sampled from a multivariate standard normal distribution.
The data is centered in the middle of the bounds used for normalization, but extends well past those bounds, resulting in larger costate values and unreasonable shooting times $\tau_s$.
Over the course of $N$ = 1,000 timesteps, the structure in the data is gradually reconstructed from the initial noise.
The dataset corresponding to the prediction of $\boldsymbol{z}_0$ after $n$ timesteps is shown for selected values of $n$.
For this visualization, $\boldsymbol{z}_0$ was calculated by plugging $\boldsymbol{z}_n$ and the predicted noise from the neural network into Eq. \eqref{eq: forward process reparameterized}. 
The displayed process here is equivalent to the visualization in Fig. \ref{fig:Diffusion Model}.
A comparison of diffusion model samples with the training data structure in Fig. \ref{fig: lamr2v1_trans_training} demonstrates that the chosen parameters produce valid results.
Slight modifications of the hyperparameters chosen in this paper may yield slightly improved results in the following sections, though are not expected to change results substantially. 
A comprehensive hyperparameter study is beyond the scope of this work. 
A more detailed explanation of the hyperparameter choices, along with results from models tested with different parameter settings for this problem, is provided in the Master's thesis of Gräbner\cite{graebner2024}.
\subsubsection{Structure in Control Predictions}
\label{section: structure in control predictions}
\begin{figure}[bt!]
    \centering
    \begin{tikzpicture}
        \node (fig1) at (0,0) {\includegraphics[width=5cm]{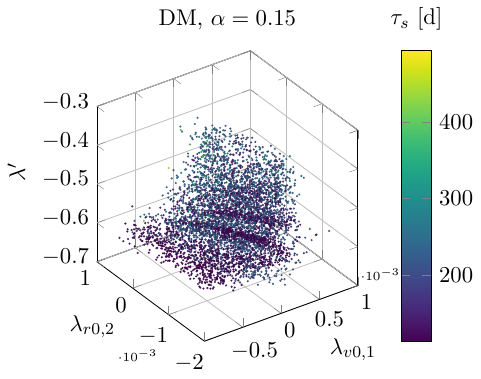}};

        \node (fig2) at (5,0) {\includegraphics[width=5cm]{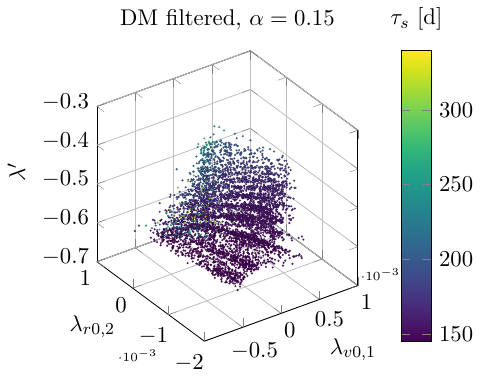}};
        
        \node (fig3) at (10, 0) {\includegraphics[width=5cm]{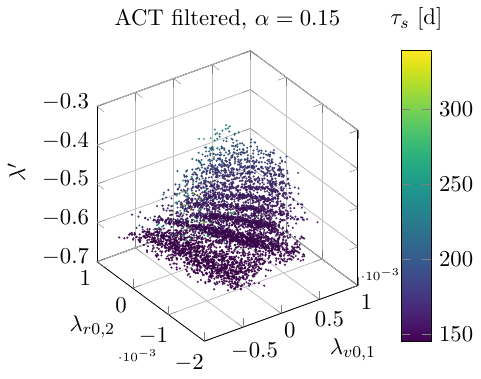}};

        \node (fig1) at (0,-4.3) {\includegraphics[width=5cm]{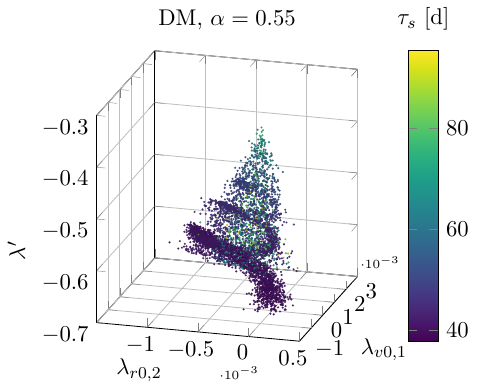}};

        \node (fig2) at (5,-4.3) {\includegraphics[width=5cm]{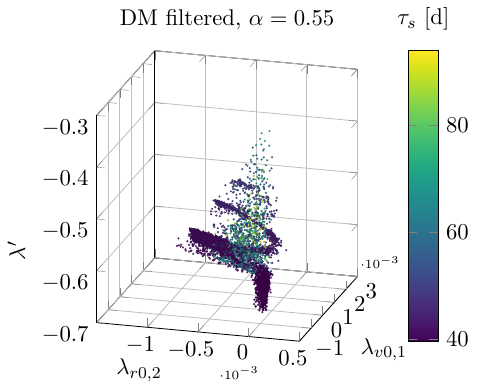}};
        
        \node (fig3) at (10, -4.3) {\includegraphics[width=5cm]{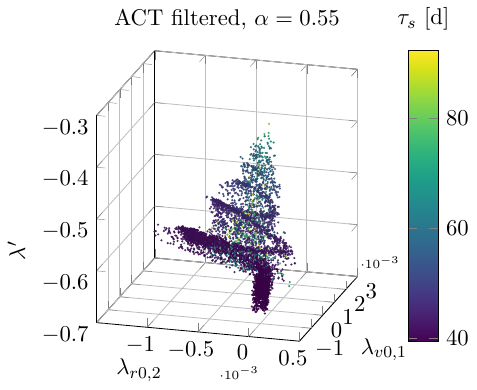}};

        \node (fig1) at (0,-8.6) {\includegraphics[width=5cm]{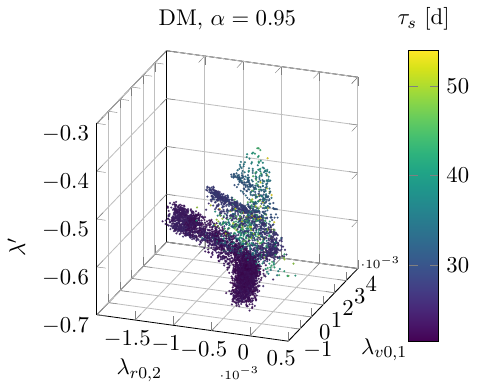}};

        \node (fig2) at (5,-8.6) {\includegraphics[width=5cm]{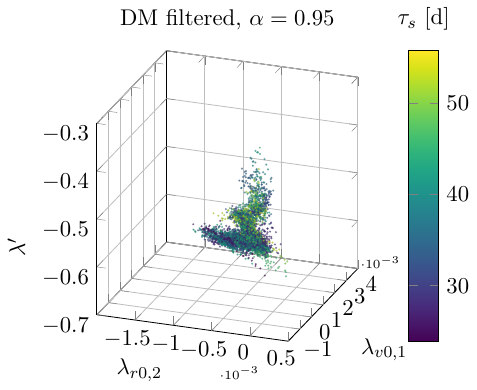}};
        
        \node (fig3) at (10, -8.6) {\includegraphics[width=5cm]{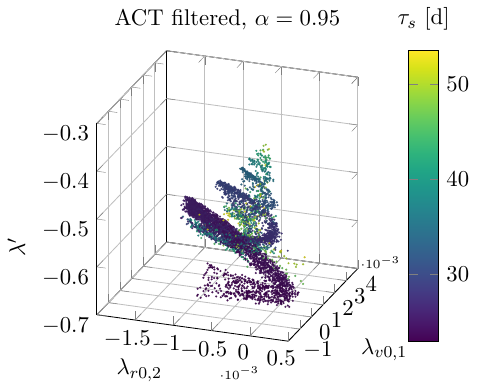}};
        
        
    \end{tikzpicture}
    \caption{Comparison of the structure in the costate space, with 5,000 data points per method for $\alpha=0.15\,(T_{max}=0.75\,\mathrm{N})$, $\alpha=0.55\,(T_{max}=2.74\,\mathrm{N})$ and $\alpha=0.95\,(T_{max}=4.73\,\mathrm{N})$.}
    \label{fig: 4D_structure_DM_TL_0.55}
\end{figure}

The diffusion model’s ability to predict changes in the structure of costate variables is demonstrated by testing it on new thrust levels that were not included in the dataset. 
We compare the generated costate samples from the model to separately generated solutions using our global search strategy.
This comparison is shown for the thrust levels $\alpha = 0.15$, $\alpha = 0.55$, and $\alpha = 0.95$ in Fig. \ref{fig: 4D_structure_DM_TL_0.55}. 
In line with the plots shown in Fig. \ref{fig: lamr2v1_trans}, the costate samples are displayed in the 3D space spanned by $\lambda_{r0,2}$, $\lambda_{v0,1}$ and the transformed costate variable $\lambda'$ from Eq. \eqref{eq: lambda_trans}.
Each scatter point corresponds to an initial costate that, when propagated forward for the shooting time indicated by the color plot, generates a candidate trajectory.
In addition to the data directly sampled from the model (which is not necessarily feasible), the filtered diffusion model data containing only feasible solutions is presented, as well as separately generated feasible solutions based on ACT initializations. 
This reference data is obtained by applying the preliminary screening algorithm after sampling from the model, filtering the solutions based on feasibility with a tolerance of $\delta = \num{1e-4}$.

For $\alpha = 0.15$, the model correctly captures the data’s location in the three-dimensional space. 
The hypersurfaces are barely visible due to a large amount of noisy data in between, which is expected since the training data for lower thrust levels is also noisy, and the ACT data for $\alpha = 0.15$ shows only a weak hypersurface structure.
For $\alpha = 0.55$, the model accurately predicts the solution structure. 
It also correctly predicts the shooting time $\tau_s$, which changes significantly across different thrust levels. 
Notably, the model successfully reproduces the trend where most of the data between the hypersurfaces exhibits longer shooting times compared to the rest of the data.
This is likely due to the solutions with longer shooting times being further away from Pareto optimality, therefore not matching the structure of most of the other solutions, which are close to the Pareto front.
The shift in the data for the thrust level $\alpha = 0.95$ is partially captured. 
While the model-generated data exhibits a similar distribution to the ACT data, it fails to reproduce part of the structure at lower $\lambda'$ values. 
Additionally, the position of the surface structure is not reproduced with sufficient accuracy.
These difficulties arise from the significant changes that occur between the two training levels, $\alpha = 0.9$ and $\alpha = 1.0$.
The separately generated data for $\alpha = 0.95$ significantly deviates from both neighboring training levels.
To achieve better results for ranges of the conditional variable where the solution structure fundamentally changes, it would be beneficial to add extra training data levels, for instance at $\alpha=0.95$.

The filtered diffusion model solutions demonstrate the model’s ability to generate valid solutions for a new problem $\mathcal{P}_{\alpha}$. 
The comparison between these model-generated solutions and the separately generated data for $\alpha = 0.15$ and $\alpha = 0.55$ shows strong alignment. 
For these thrust levels, the results also highlight the model’s capacity to create diverse solutions that are not concentrated in a single large data cluster.
For $\alpha = 0.95$, the filtered data exhibits a distribution that significantly differs from both the unfiltered diffusion model data and the ACT data. 
Since the model’s predictions slightly misalign with the actual hypersurface locations, most of the samples do not produce feasible solutions. 
Consequently, a majority of the filtered solutions originate from the noisier data samples with longer shooting times. 
This leads to poorer model performance compared to $\alpha = 0.15$ and $\alpha = 0.55$, as confirmed by the results in the following section.
\subsubsection{Model Performance}
\label{section: warmstart performance}
An ablation study for $\alpha = 0.15$, $\alpha = 0.55$, and $\alpha = 0.95$ in Table \ref{tab: europa comp NO} showcases the model's capability to produce new, high-quality solutions at a higher rate than the ACT method.
\begin{table}[b!]
	\centering
	\caption{Diffusion model sampling results compared to samples from the ACT method for 300,000 initializations per thrust level and a feasibility tolerance of $10^{-4}$. 
    Solutions/minute include both sampling and solving times.}
		\begin{tabular}{lcccccc} \toprule
          & \multicolumn{2}{c}{$\alpha=0.15\,(T_{max}=0.75\,\mathrm{N})$} & \multicolumn{2}{c}{$\alpha=0.55\,(T_{max}=2.74\,\mathrm{N})$} & \multicolumn{2}{c}{$\alpha=0.95\,(T_{max}=4.73\,\mathrm{N})$}\\ 
           \cmidrule(lr){2-3} \cmidrule(lr){4-5} \cmidrule(lr){6-7}
		   & \textbf{ACT} & \textbf{DM} & \textbf{ACT} & \textbf{DM} & \textbf{ACT} & \textbf{DM} \\ 
       \midrule
		\textbf{Feasibility ratio}  [\%] & 0.065 & $\mathbf{7.58}$ & 0.035 & $\mathbf{9.79}$ & 0.025 & $\mathbf{4.21}$ \\
        \textbf{Runtime [min]} & $\mathbf{164.15}$ & 884.03 & $\mathbf{48.28}$ & 257.61 & $\mathbf{32.95}$ & 52.35 \\   
        \textbf{Sampling time [min]} & $\mathbf{0.0}$ & 20.20 & $\mathbf{0.0}$ & 20.20 & $\mathbf{0.0}$ & 20.18 \\
        \textbf{Solutions/min.} & 1.18 & $\mathbf{25.14}$ & 2.17 & $\mathbf{105.75}$ & 2.28 & $\mathbf{174.25}$ \\
         \textbf{Average $\Delta v\,[\unit{m.s^{-1}}]$} & $\mathbf{349.82}$ & 358.24& $\mathbf{360.40}$& 364.72& 358.24& $\mathbf{349.82}$\\
        \bottomrule
	\end{tabular}
	\label{tab: europa comp NO}%
\end{table}
For this study, the preliminary screening algorithm was used with a convergence tolerance of $\delta = \num{1e-4}$, corresponding to the global search strategy represented by the lower path in Fig. \ref{fig: global search}. 
For the ACT method, the feasibility ratio remains close to $0.0\,\%$ across all three thrust values. 
Despite this, solutions are generated at a high rate of more than one per minute due to the low runtime of the ACT method. 
This highlights the strength of the preliminary screening algorithm which evaluates the feasibility of a solution in only a fraction of a second, allowing thousands of solutions to be processed in a short period.
Due to the high quality of the diffusion model samples, the feasibility ratios are increased by more than two orders of magnitude.
This is remarkable, considering that the model itself has no direct knowledge of the dynamics and constraints of the transfer. 
The feasibility ratio for $\alpha=0.95$ is lower than the values for $\alpha=0.15$ and $\alpha=0.55$.
This aligns with the results from the previous section, where the costate structure for $\alpha = 0.95$ was not captured as accurately. 
Nonetheless, the feasibility ratio remains significantly higher compared to samples from the ACT method.

\begin{figure}[b!]
    \centering
    \begin{subfigure}[b]{0.34\textwidth}
        \centering
        \includegraphics[scale=0.65]{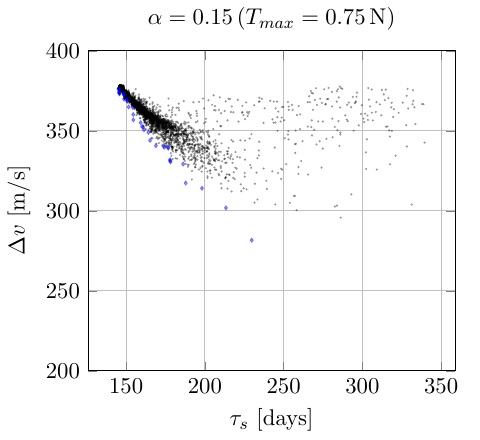}
    \end{subfigure}\hfill
    \begin{subfigure}[b]{0.33\textwidth}
        \centering
        \includegraphics[scale=0.65]{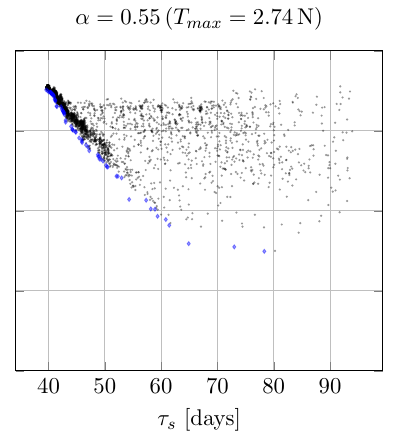}
    \end{subfigure}\hfill
        \begin{subfigure}[b]{0.33\textwidth}
        \centering
        \includegraphics[scale=0.65]{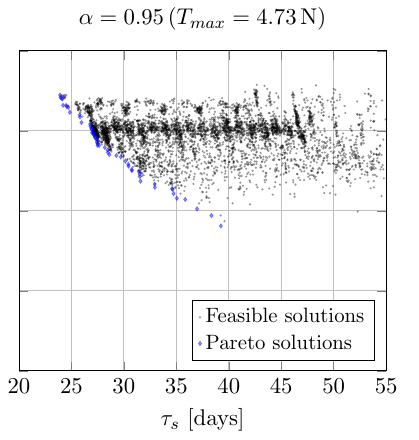}
    \end{subfigure}\hfill
    \caption{A subset of 5,000 solutions generated from diffusion model samples for the three testing thrust levels, displayed in the $\Delta v$-$\tau_s$ plane. A pareto front is uncovered in all three cases.}
    \label{fig: dv-TOF plots DM}
\end{figure}

Sampling from the diffusion model enables solution generation at a high rate for the test problems.
Although the runtime is longer compared to the ACT method, and additional GPU sampling time must be accounted for, the number of solutions generated per minute is significantly higher with diffusion model samples.
For both methods, runtime decreases as the thrust level increases, which is due to shorter shooting times and faster trajectory propagation in the preliminary screening algorithm.
The higher runtimes for diffusion model samples compared to the ACT method are a result of the greater number of feasible solutions found, which requires additional operations to save the results. 
This time could be further reduced by optimizing the code for these operations and only saving the essential parameters of each solution.
The performance improvement is most notable for $\alpha = 0.95$, with a factor of more than 75, while for the more challenging problem of $\alpha = 0.15$, the solutions per minute still increase by a factor of over 20.
The solutions generated by the diffusion model are of high quality with respect to the objective, as evidenced by a low average $\Delta v$. 
Since the Pareto front was explicitly targeted during the training data generation, the solutions produced by the diffusion model are also close to Pareto optimality. 
This is further illustrated in Fig. \ref{fig: dv-TOF plots DM}, where the Pareto front for the diffusion model solutions is visualized.

\newpage
\subsection{GTO Halo Transfer}
\label{sec:GTOHalo}
\subsubsection{Problem Description}
\label{sec: problem description GTO}
\begin{figure}[b!]
    \centering
    \includegraphics[scale=0.8]{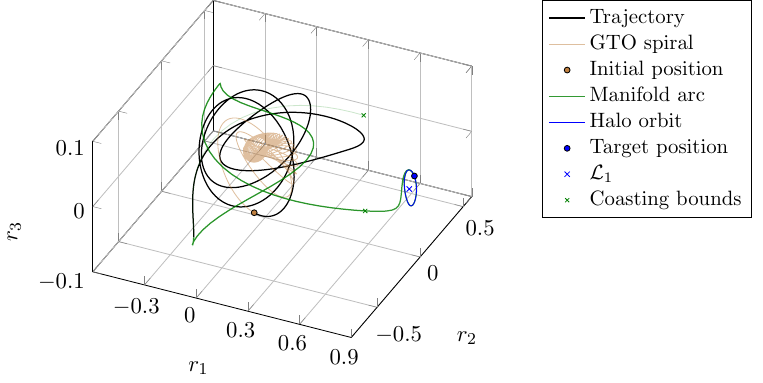}
    \caption{Example trajectory with $T_{max} = 0.5\,\unit{N}$ ($\alpha=0.0$) for the GTO halo transfer in the CR3BP frame. 
    }
    \label{fig:example trajectory GTO}
\end{figure}

    \begin{table}[b!]
    \centering
    \caption{Problem parameters for GTO Halo transfer.}
    \small 
    \setlength{\tabcolsep}{6pt} 
    \begin{tabular}{lclc} 
    \toprule
    \multicolumn{2}{l}{\textbf{Trajectory parameters}} & \multicolumn{2}{c}{} \\
    \midrule
    \multicolumn{2}{l}{Initial state $[\boldsymbol{r}_0^T,\boldsymbol{v}_0^T]$ [NU]} & \multicolumn{2}{c}{$[ 0.0763, -0.2786, -0.0445, 0.9971, 0.7897, -0.2632]$}\\
    \multicolumn{2}{l}{Specific energy initial state [NU]} & \multicolumn{2}{c}{$-2.5517$}\\
    \multicolumn{2}{l}{Terminal state $[\boldsymbol{r}_f^T,\boldsymbol{v}_f^T]$ [NU]} & \multicolumn{2}{c}{$[0.8380, 0.0579, 0.0110, 0.0427, 0.0464, -0.0507]$}\\
    \multicolumn{2}{l}{Specific energy halo orbit [NU]} &\multicolumn{2}{c}{$-1.5842$}\\
    \multicolumn{2}{l}{Final coast range $\tau_f$ [TU]} & \multicolumn{2}{c}{$[5.0,11.0]$}\\
    \midrule
    \multicolumn{2}{l}{\textbf{Spacecraft parameters}} & \multicolumn{2}{l}{\textbf{Natural units (Earth-Moon)}} \\
    \midrule
    Initial mass $m_0$ [kg] & 1,000 & Distance unit [km] & $384,400$ \\
    Fuel mass [kg] & $700$ & Time unit [s] & $375,201.92$ \\
    Dry mass [kg] & $300$ & Mass unit [kg] & $6.046 \times 10^{24}$ \\
    Specific impulse $I_{sp}$ [s] & 1,000 & & \\
    Thrust range $T_{max}$ [N] & $[0.5, 1.0]$ & & \\
    \bottomrule
    \label{tab:problem parameters GTO Halo}
    \end{tabular}
    \end{table}

The mission optimized in this problem is similar to the problem solved by Li et al. \cite{Li.872023} and involves a transfer from an Earth orbit to a halo orbit around libration point $\mathcal{L}_1$. 
An example trajectory is shown in Fig. \ref{fig:example trajectory GTO}. 
Starting from the initial geostationary transfer orbit (GTO), the orbital energy is increased through a fixed-time tangential low-thrust spiral, which is only computed once and is identical for all solutions.
The endpoint of this spiral serves as the initial boundary condition for the trajectory optimization problem.
The dynamical structures of the CR3BP system are actively incorporated into the trajectory design by targeting the halo orbit through a final coast, $\tau_f$, along its stable interior invariant manifold. 
The terminal boundary condition is defined as any point on a fixed arc of this manifold, within the coasting bounds marked in Fig. \ref{fig:example trajectory GTO}.
The specific point on the manifold arc that serves as the boundary condition is determined by the preliminary screening algorithm and parameterized through the final coast time.
The detailed problem parameters are provided in Table \ref{tab:problem parameters GTO Halo}. 
The bounds on the final coast time correspond to the time required to coast from the coasting bounds, marked in Fig. \ref{fig:example trajectory GTO}, to the target position on the halo orbit. 
The target position is the seed point of an arbitrarily chosen manifold arc on the halo orbit with a specified orbital energy.
The spacecraft parameters are consistent with those used in Li et al. \cite{Li.872023}. 
The maximum thrust, $T_{max}$, is treated as the conditional problem parameter and is varied within a specified range. 
The constants used to normalize the distance, time, and mass units to the system of natural units in this section are also listed in Table \ref{tab:problem parameters GTO Halo}. 
Unless otherwise specified, all variables in this section are normalized using these natural units.
Solving this problem presents a greater challenge than the Europa DRO transfer described earlier, as it involves a three-dimensional transfer with a wide range of qualitatively different solutions and a dominant forcing component provided by the primary. 

\subsubsection{Data Generation}
\label{sec: data generation GTO}
\begin{table}[b!]
\centering
\caption{Global search parameters for the GTO halo transfer. The chosen bounds ensure that $S>\lambda_{m}m/c$ at $t=0$, with $\lambda_{m,0}m_0/c=-0.10447$.}
\setlength{\tabcolsep}{8pt} 
\begin{tabular}{p{4.8cm}c|p{4cm}c} 
\toprule
\textbf{Parameter} & \textbf{Value/Range} & \textbf{Parameter} & \textbf{Value/Range} \\
\toprule
Global search tolerance $\delta$ [NU] & $0.05$ & SNOPT runtime limit [s] & $30$\\
Feasibility tolerance [NU] & $0.001$ & SNOPT mode & optimal \\
$\varphi_0$ [rad] & $[-0.1,\,1.25]$ & $\dot{\varphi}_0$ [rad/TU] & $[-15.0,\,0.0]$ \\
$\beta_0$ & $[0.4,\,1.3]$ & $\dot{\beta}_0$ & $[-6.0,\,-1.0]$ \\
$S_0-\lambda_{m,0}m_0/c$ [NU] & $[0.08,\,0.4]$ & $\dot{S}_0$ [NU] & $[-0.5,\,0.9]$ \\
\bottomrule
\end{tabular}
\label{tab: GS parameters GTO Halo}
\end{table}
The conditional parameter $\alpha$ for this problem represents the maximum thrust magnitude of the spacecraft,
with $\alpha = (T_{max}-0.5\,\mathrm{N})/(1.0\,\mathrm{N}-0.5\,\mathrm{N})$, where $\alpha\in[0.0,1.0]$.
For diffusion model training, a dataset consisting of solutions for eleven fixed thrust levels ranging from $\alpha=0.0$ to $\alpha=1.0$ in increments of $0.1$ is created.
The dataset consists of 297,000 solutions, with 27,000 solutions per thrust level. 
The global search parameters used for data generation are listed in Table \ref{tab: GS parameters GTO Halo}.
Given the increased complexity of this problem, the upper path of the global search strategy from Fig. \ref{fig: global search} was used for data generation. 
Relying solely on the preliminary screening algorithm without numerical optimization proved inefficient for this three-dimensional low-thrust transfer.
For actual mission design, a tighter local convergence tolerance than the value specified in Table \ref{tab: GS parameters GTO Halo} would be required, however the selected tolerance is sufficient to capture the structure of the solution space. 
In general it is also no problem to achieve a lower convergence tolerance by warm-starting the solver and allowing for additional runtime.

The ranges specified for $\varphi_0$, $\dot{\varphi}_0$, $\beta_0$, $\dot{\beta}_0$, $S_0$, and $\dot{S}_0$ represent the maximum bounds applied across all thrust levels.
Since the orbital energy increases from the initial position at the end of the GTO spiral to the target halo orbit, the thrust angles $\varphi_0$ and $\beta_0$ are initialized so that the resulting thrust vector has a positive component along the flight direction. 
The additional components perpendicular to the velocity vector resulting from the specific values of $\varphi_0$ and $\beta_0$, are more challenging to interpret intuitively. 
The same applies to $\dot{\varphi}_0$ and $\dot{\beta}_0$, even though it can be observed that with the angles being initialized with mostly positive values, their derivatives are negative, moving the thrust vector back towards the flight direction.
For the majority of initializations, the switching function $S_0$ is positive, indicating a thrusting phase at the beginning of the transfer.

To determine these ranges, the problem was initially solved using broader intervals, which were iteratively refined to the region where most solutions are concentrated for each thrust level. 
Although the control variables are more intuitive than the position and velocity costates, identifying suitable ranges - particularly for the derivatives - remains challenging. 
This is one of the limitations of the ACT method, necessitating multiple extensive test runs to establish reasonable bounds. Additionally, the ideal ranges vary for each thrust level, further complicating the process.  
\begin{figure}[t!]
    \centering
    \begin{subfigure}[b]{0.35\textwidth}
        \centering
        \includegraphics[scale=0.66]{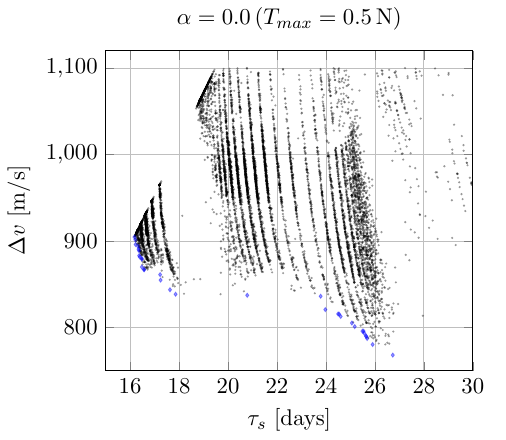}
    \end{subfigure}\hfill
        \begin{subfigure}[b]{0.32\textwidth}
        \centering
        \includegraphics[scale=0.66]{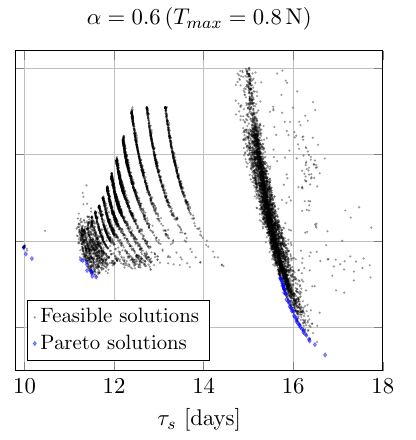}
    \end{subfigure}\hfill
    \begin{subfigure}[b]{0.32\textwidth}
        \centering
        \includegraphics[scale=0.66]{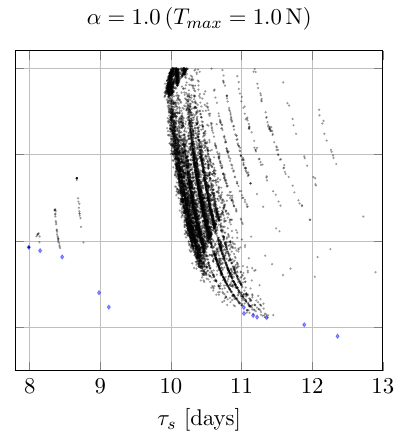}
    \end{subfigure}
    \caption{A subset of 5,000 solutions from the training dataset of the GTO halo transfer for each of three thrust levels $\alpha$, displayed in the $\Delta v$-$\tau_s$ plane.}
    \label{fig: dv-TOF plots GTO}
\end{figure}

While this approach enables efficient generation of solutions, it does not ensure a thorough exploration of the solution space. 
This limitation of the ACT method is evident in Fig. \ref{fig: dv-TOF plots GTO}, which visualizes a subset of the training dataset for three different thrust levels in the frame of the competing objectives $\Delta v$ and shooting time $\tau_s$.
The training data has been filtered to include only trajectories with $\Delta v \le \text{1,100}\,\unit{m.s^{-1}}$.
Note that the $\Delta v$ presented here does not include the velocity change required for the GTO low-thrust spiral.
The spiral is fixed for all simulations, and hence that portion has an identical $\Delta v$ for all solutions with a value of $2,082\,\unit{m.s^{-1}}$.
The solutions for all thrust levels fall within a similar range of $\Delta v$, whereas the range of the shooting time decreases for higher thrust levels.  
Clusters of solutions can be observed for each thrust level, and it remains uncertain whether the gaps between these clusters are due to insufficient exploration of the search space or a lack of feasible solutions in those regions. 
The points labeled as Pareto solutions in the plots are Pareto optimal within the training dataset, but it is not guaranteed that there are no other solutions with both lower $\Delta v$ and $\tau_s$.
Nonetheless, this dataset comprises a set of feasible and diverse solutions, which are suitable to demonstrate the capabilities of the generative learning framework, without claiming that the generated samples are close to globally optimal solutions. 
\subsubsection{Solution Structure}
The structure of the six-dimensional costate space is examined by applying a Principal Component Analysis (PCA) to the training dataset. 
PCA is a a linear dimensionality-reduction method that orthogonally transforms the data into principal components ranked by the variance they explain \cite{Jolliffe2011}.
We project the data onto the space defined by the first three principal components, labeled $PC1$, $PC2$, and $PC3$. 
The explained variance ratios for these components are $59.98\,\%$ for $PC1$, $27.01\,\%$ for $PC2$, and $12.50\,\%$ for $PC3$. 
Together, these three principal components account for over $99\,\%$ of the variance in the data, effectively capturing the vast majority of the information. 
The costate vectors are mapped to this new three-dimensional space using the transformation:
\begin{equation*}
\begin{pmatrix}
PC1 \\
PC2 \\
PC3 \\
\end{pmatrix}
=
\begin{pmatrix}
0.408 & -0.853 & -0.289 & 0.117 & 0.052 & -0.074 \\
0.588 & 0.486  & -0.564 & -0.031 & -0.126 & -0.289 \\
-0.654 & -0.068 & -0.744 & -0.094 & 0.074  & -0.014 \\
\end{pmatrix}
\left(
\begin{pmatrix}
\lambda_{r1} \\
\lambda_{r2} \\
\lambda_{r3} \\
\lambda_{v1} \\
\lambda_{v2} \\
\lambda_{v3} \\
\end{pmatrix}
-
\begin{pmatrix}
0.076 \\
0.927 \\
-0.464 \\
-0.096 \\
-0.114 \\
-0.182 \\
\end{pmatrix}
\right),
\end{equation*}
where the costates are first centered around the mean and then multiplied with the transformation matrix.
\label{sec: solution structure GTO}
\begin{figure}[t!]
    \centering
    \begin{subfigure}[b]{0.33\textwidth}
        \centering
        \includegraphics[scale=0.68]{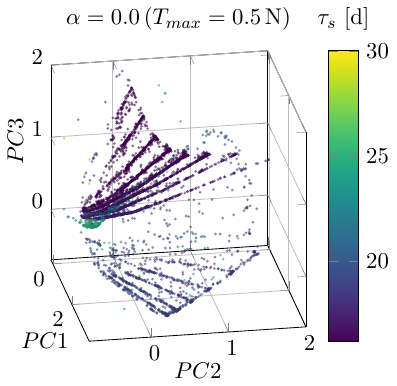}
    \end{subfigure}\hfill
    \begin{subfigure}[b]{0.33\textwidth}
        \centering
        \includegraphics[scale=0.68]{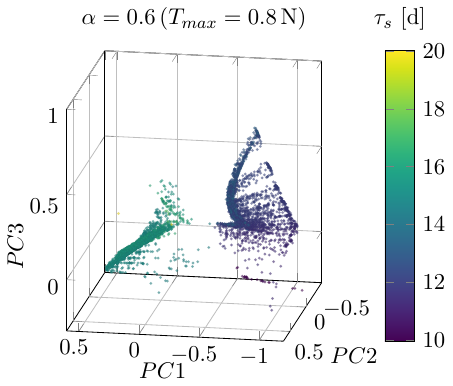}
    \end{subfigure}\hfill
   \begin{subfigure}[b]{0.33\textwidth}
        \centering
        \includegraphics[scale=0.68]{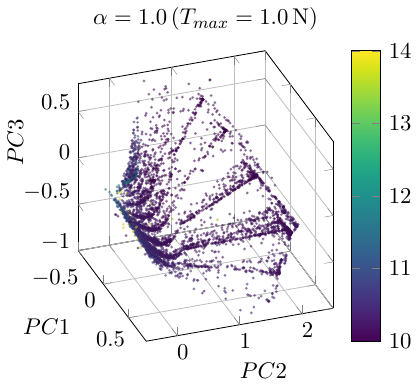}
    \end{subfigure}
    \caption{Structure in the principal component space of the GTO halo transfer for three different thrust levels $\alpha$. Each plots displays a subset of 5,000 solutions from the training data.}
    \label{fig: PCA_sol_struc}
\end{figure}

Visualizing the training dataset in the frame of the first three principal components reveals a complex structure. 
As shown in Fig. \ref{fig: PCA_sol_struc} for three different thrust levels, clusters of solutions emerge in the principal component space, with their locations and shapes varying for each thrust level. 
The different axis ranges in the plots further emphasize the significant shifts in the costate space, when varying the maximum thrust magnitude. 
For $\alpha=0.6$, the data groups into two primary clusters, which are associated with different ranges of the shooting time $\tau_s$, as indicated by the colormap in Fig. \ref{fig: PCA_sol_struc}.
A similar behaviour can be observed for $\alpha=0.0$.
The stripes of solutions within the clusters can be observed in the $\Delta v$-$\tau_s$ plane in Fig. \ref{fig: dv-TOF plots GTO} and are also present in the principal component space in Fig. \ref{fig: PCA_sol_struc}.
The presence of clusters in the costate space reinforces the rationale for applying generative learning methods to this problem. 
With major changes occurring in the costate space as the thrust level changes, generating data for new thrust levels is a complex task.
This presents a significant challenge for the diffusion model, which is tasked with generating new solutions for different values $\alpha$ in the following sections. 
\subsubsection{Structure in Control Predictions}
\label{sec: structure in control predictions GTO}
During training, the diffusion model learns the probability distribution of the costate space conditioned on the thrust level. 
We validate the model’s ability to generalize this distribution to new problems by testing it on the thrust levels $\alpha = 0.05$,   $\alpha = 0.45$ and  $\alpha = 0.95$.

\begin{figure}[t!]
    \centering
    \begin{tikzpicture}
        \node (fig1) at (0,0) {\includegraphics[scale=0.55]{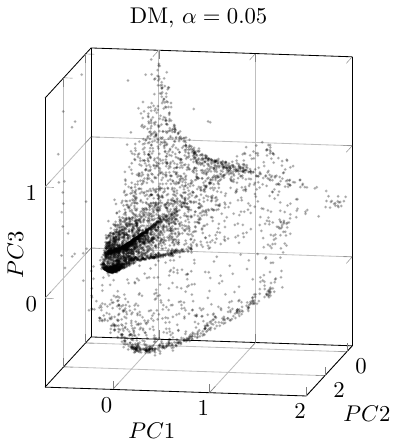}};

        \node (fig2) at (4.7,0) {\includegraphics[scale=0.55]{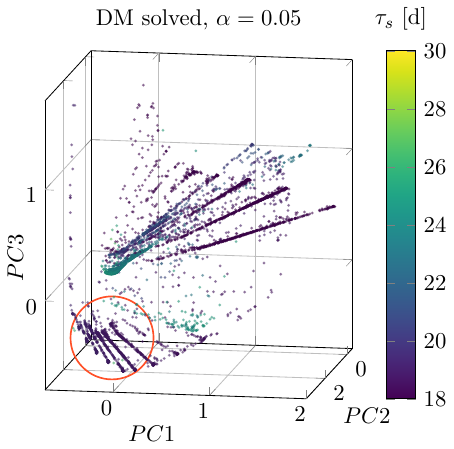}};
        
        \node (fig3) at (10.2, 0) {\includegraphics[scale=0.55]{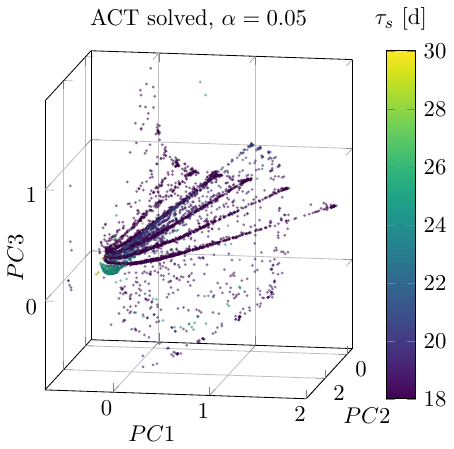}};

        \node (fig1) at (0,-4.3) {\includegraphics[scale=0.55]{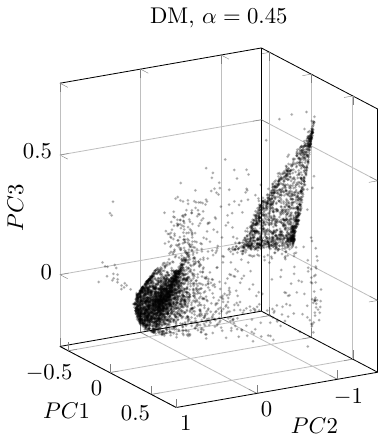}};

        \node (fig2) at (4.7,-4.3) {\includegraphics[scale=0.55]{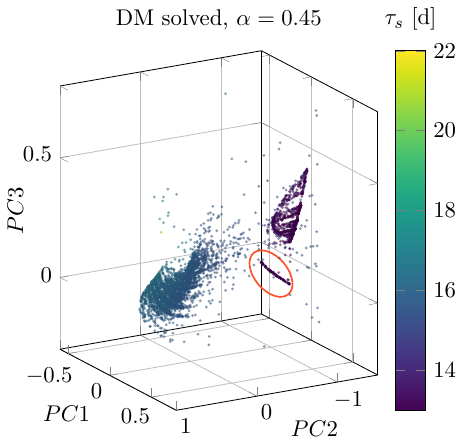}};
        
        \node (fig3) at (10.2, -4.3) {\includegraphics[scale=0.55]{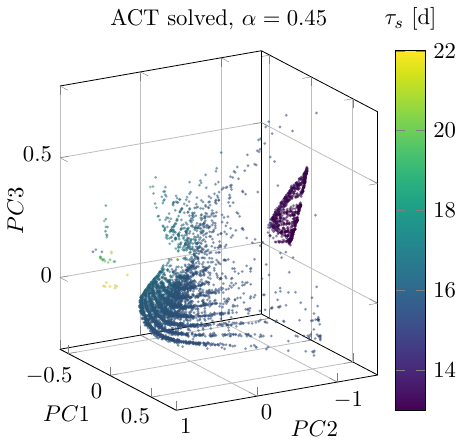}};

        \node (fig1) at (0,-8.6) {\includegraphics[scale=0.55]{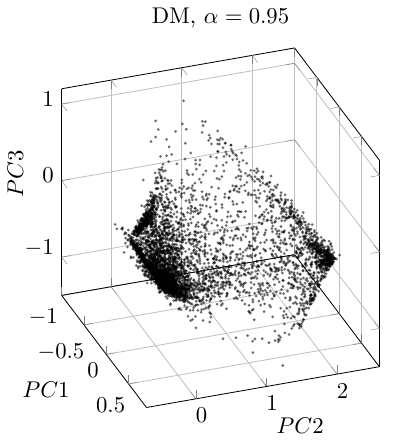}};

        \node (fig2) at (4.7,-8.6) {\includegraphics[scale=0.55]{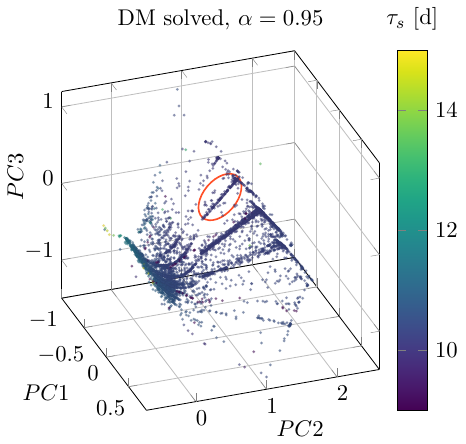}};
        
        \node (fig3) at (10.2, -8.6) {\includegraphics[scale=0.55]{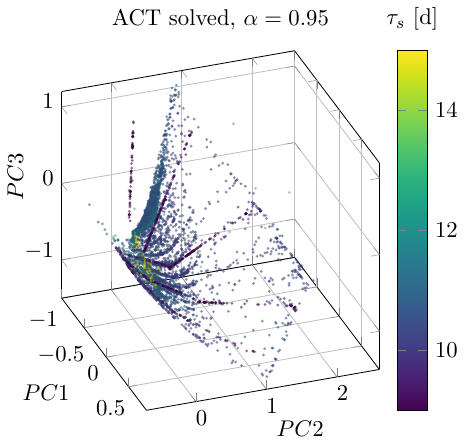}};
        
        
    \end{tikzpicture}
    \caption{Comparison of the structure in the costate space, with 5,000 data points per method. Results are shown for thrust levels $\alpha=0.05\,(T_{max}=0.525\,\mathrm{N})$, $\alpha=0.45\,(T_{max}=0.725\,\mathrm{N})$ and $\alpha=0.95\,(T_{max}=0.975\,\mathrm{N})$.}
    \label{fig: PCA_TL_0.45}
\end{figure}

The diffusion model samples are transformed into the principal component space and compared with separately generated data from the ACT method.
For all three tested thrust levels, the samples generated by the diffusion model closely resemble the separately generated data. 
This similarity is illustrated for 5,000 data points per thrust level and method in Fig. \ref{fig: PCA_TL_0.45}.
Additionally, each figure includes the final set of feasible solutions obtained by warm-starting SNOPT with the diffusion model samples. 
The shooting time of feasible solutions from both the ACT method and the diffusion model is visualized using a color plot, whereas the diffusion model samples, which do not include that information, are shown in black.
The model accurately captures how the structure in the principal component space shifts as the thrust level increases, with samples being generated within all main solution clusters. 
The final solutions generated from the diffusion model samples exhibit a structure that closely mirrors the ACT data for all shown thrust levels, including similar time of flight values. 
While the shape of some clusters slightly deviates in the diffusion model samples, leading to the model missing some solutions, it also generates solutions in regions not targeted by the ACT method.
Some of these regions are highlighted in orange in Fig. \ref{fig: PCA_TL_0.45} 


\subsubsection{Model Performance}
\label{sec: warmstart performance GTO}
By using the diffusion model samples, visualized in the previous section as initial guesses for the numerical solver SNOPT, solutions for a new problem $\mathcal{P}_{\alpha}$ are generated efficiently. 
\begin{figure}[b!]
    \centering
    \begin{subfigure}[t]{0.245\textwidth}
        \centering
        \vspace{0pt}
        \includegraphics[scale=0.75]{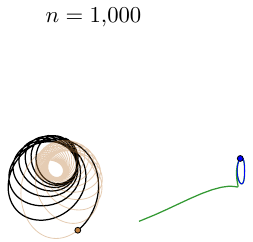}
    \end{subfigure}\hfill
    \begin{subfigure}[t]{0.245\textwidth}
        \centering
        \vspace{0pt}
        \includegraphics[scale=0.75]{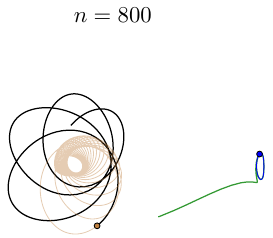}
    \end{subfigure}\hfill
    \begin{subfigure}[t]{0.245\textwidth}
        \centering
        \vspace{0pt}
        \includegraphics[scale=0.75]{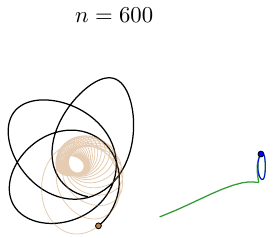}
    \end{subfigure}\hfill
    \vspace{0.7cm}
    \begin{subfigure}[t]{0.245\textwidth}
        \centering
        \vspace{0pt}
        \includegraphics[scale=0.75]{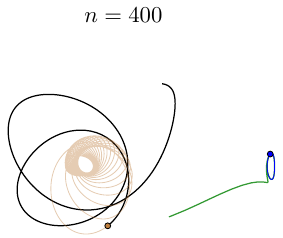}
    \end{subfigure}
    \begin{subfigure}[t]{0.245\textwidth}
        \centering
        \includegraphics[scale=0.75]{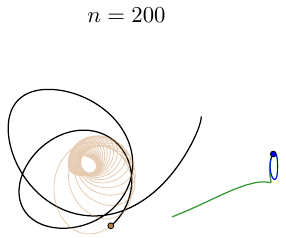}
    \end{subfigure}\hfill
    \vspace{0.7cm}
    \begin{subfigure}[t]{0.245\textwidth}
        \centering
        \includegraphics[scale=0.75]{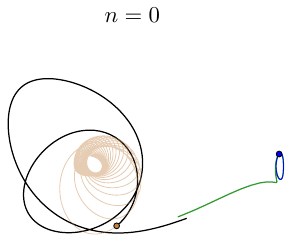}
    \end{subfigure}
    \begin{subfigure}[t]{0.49\textwidth}
        \raggedleft
        \includegraphics[scale=0.75]{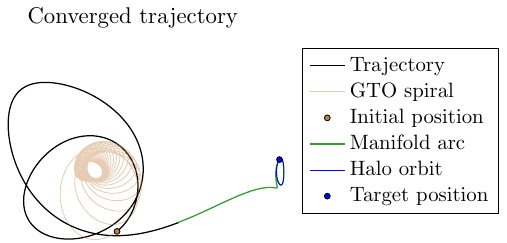}
    \end{subfigure}\hfill
    \begin{tikzpicture}[overlay, remember picture]
        \draw[->, thick] (-7.7,1.3) -- node[above, font=\small] {SNOPT} (-6.5,1.3);
    \end{tikzpicture}
    \vspace{0.7cm}
    \caption{Trajectories corresponding to control vectors from different diffusion timesteps $n$ for the GTO halo transfer with $\alpha=0.45\,(T_{max}=0.725\,\mathrm{N})$.} 
    \label{fig: timestep trajectories GTO}
\end{figure}
The process by which the diffusion model generates a high-quality initial guess for the costate vector, which allows to quickly solve for a feasible trajectory is visualized in Fig. \ref{fig: timestep trajectories GTO}.
The model begins with noisy data sampled from a multivariate standard normal distribution and iteratively reconstructs structure in the data through a series of timesteps guided by the pre-trained neural network.
Each trajectory in Fig. \ref{fig: timestep trajectories GTO}, corresponds to the trajectory resulting from the prediction for $\boldsymbol{z}_0$ after $(N-n)$ diffusion timesteps.
To stay consistent with the reverse diffusion process visualized in Fig. \ref{fig:Diffusion Model}, the timestep label decreases from $n=N$ to $n=0$.
The displayed trajectories are generated by propagating the initial state and initial costate guess forward in time.
For consistency, the shooting time $\tau_s$ and final coast time $\tau_f$ for all displayed trajectories are set according to the values from the final converged trajectory.
The first guess at $n=1$,000 results in an unreasonable trajectory that spirals into the Earth. 
As $n$ decreases, the model's predictions become more reasonable and the magnitude of the constraint violations at the trajectory endpoint decreases. 
The actual sample at $n=0$ provides a good approximation of the solution.
This guess is then used to warm-start the numerical solver SNOPT, which requires less than $10\,\mathrm{s}$ to converge to a feasible trajectory. 

Although this process is visualized for just one example case, the overall high quality of diffusion model samples is confirmed across the testing thrust levels $\alpha=0.05$, $\alpha=0.45$ and $\alpha=0.95$, by comparing it to initial guesses sampled from a uniform distribution (referred to as uniformly sampled - US) and samples from the ACT method in Table \ref{tab: DM_comp_GTO}.
For each method, the results are based on 100,000 samples per thrust level, which are first filtered using the preliminary screening algorithm with a global search tolerance of $\delta=0.05$ and then solved with a maximum SNOPT runtime of $30\,\si{s}$ and a feasibility tolerance of $10^{-3}$.
We perform these tests with the same coarse tolerances used for the training data and provide a study including tighter tolerances in the subsequent section.

The costate ranges for uniform sampling were derived from the maximum and minimum values found in the training datasets for the two neighboring thrust levels.
The diffusion model demonstrates a significantly higher feasibility ratio than the two other methods, surpassing them by two orders of magnitude across all three thrust values. 
This is primarily due to the higher ratio of local to global attempts, meaning a greater percentage of diffusion model samples already represents an approximate solution within the global search tolerance $\delta$. 
Furthermore, a larger number of samples from the diffusion model that pass the preliminary screening algorithm successfully converge to feasible solutions, as evidenced by the higher ratio of converged to local attempts.
The increased number of local attempts contributes to a longer runtime for the diffusion model samples.
However, despite this, the solutions per minute produced by the diffusion model still outperform both the US and ACT methods by more than an order of magnitude.
We also obtain more solutions per minute than Li et al. \cite{li2024diffusolvediffusionbasedsolvernonconvex}, who applied a diffusion model framework to a similar problem using the direct method.
Regarding the average $\Delta v$ values, there is no consistent trend across the methods, with each method delivering the best results for different thrust values.
All the $\Delta v$ averages are within a similar range, except for the US method at $\alpha=0.45$, where the average $\Delta v$ is noticeably higher.
\begin{table}[t!]
    \centering
    \caption{Comparison of solver warm-start performance for 100,000 initializations per method and a feasibility tolerance of $10^{-3}$. Solutions per minute include both solving and sampling time for the diffusion model.}
    \setlength{\tabcolsep}{3pt} 
    \begin{tabular}{lcccccccccc}
        \toprule
         & \multicolumn{3}{c}{$\alpha = 0.05\,(T_{max}=0.525\,\mathrm{N})$} & \multicolumn{3}{c}{$\alpha = 0.45\,(T_{max}=0.725\,\mathrm{N})$} & \multicolumn{3}{c}{$\alpha = 0.95\,(T_{max}=0.975\,\mathrm{N})$} \\
        \cmidrule(lr){2-4} \cmidrule(lr){5-7} \cmidrule(lr){8-10}
        & \textbf{US} & \textbf{ACT} & \textbf{DM} & \textbf{US} & \textbf{ACT} & \textbf{DM} & \textbf{US} & \textbf{ACT} & \textbf{DM} \\
        \midrule
        \textbf{Local/global} [\%] & 0.25 & 0.36 & $\mathbf{16.12}$ & 0.11 & 0.25 & $\mathbf{74.43}$ & 0.69 & 0.63 & $\mathbf{92.08}$ \\
        \textbf{Converged/local} [\%] & 18.50 & 12.47 & $\mathbf{21.90}$ & 8.49 & 14.00 & $\mathbf{27.53}$ & 19.19 & 35.41 & $\mathbf{67.50}$ \\
        \textbf{Feasibility ratio} [\%] & 0.047 & 0.045 & $\mathbf{3.53}$ & 0.009 & 0.035 & $\mathbf{20.49}$ & 0.13 & 0.22 & $\mathbf{62.15}$ \\ 
        \textbf{Runtime} [min.] & 493 & $\mathbf{309}$ & 2123 & 334 & $\mathbf{294}$ & 7774 & 248 & $\mathbf{227}$ & 4341 \\ 
        \textbf{Solutions/min.} & 0.095 & 0.15 & $\mathbf{1.66}$ & 0.027 & 0.12 & $\mathbf{2.64}$ & 0.54 & 0.98 & $\mathbf{14.32}$ \\
        \textbf{Average $\Delta v$ [m/s]} & 1024.33 & $\mathbf{1000.76}$ & 1052.13 & 1244 & 973.16 & $\mathbf{950.92}$ & $\mathbf{955.29}$ & 989.17 & 1019.20\\ 
        \bottomrule
    \end{tabular}
    \label{tab: DM_comp_GTO}
\end{table}

\subsubsection{Tolerance Refinement}
\label{sec: tolerance refinement}
A necessary property of any useful solutions found from a global search using low-fidelity physics models, as is done in this paper, is that the solutions can be continued into high-fidelity models with tight feasibility tolerances. 
In the global search phase, where computational speed is important, the feasibility tolerances can be looser as long as they can be tightened at a later time. 
The feasibility tolerances used in this study ($10^{-4}$ NU for the Europa DRO transfer and $10^{-3}$ NU for the GTO Halo transfer) are coarse by mission‐design standards. 
As the aim of the paper is to demonstrate the ability to learn the global control structure using generative machine learning given a subset of data solved at a specified tolerance, we should expect that the trained model will provide control predictions that satisfy this specified tolerance and may struggle to provide as many predictions that satisfy a tighter tolerance. 

To understand how robust our architecture was to providing samples that could be solved to a tighter tolerance, 
we perform a study for the GTO Halo transfer case that examines the percentage of solutions that remain feasible for ever tighter feasibility tolerances. 
The GTO Halo transfer was used because it is dynamically more complex and was originally solved with the loosest state tolerance. 
Table~\ref{tab: tol_refinement} reports the convergence rate when SNOPT is warm-started with diffusion-model samples that were feasible at $10^{-3}$ NU. 
As the tolerance tightens, fewer runs converge within the $30\,\mathrm{s}$ time limit.
Nonetheless, more than two-thirds still converge at $10^{-6}$, which corresponds to a maximum position error of $385\,\mathrm{m }$ and a maximum velocity error of $1.025\times10^{-3}\unit{m.s^{-1}}$.
These values are adequate for preliminary CR3BP analyses, as higher-fidelity studies would require a more detailed dynamical model and the additional fidelity will have a more dramatic effect on the ability to continue these solutions versus the state error due to the solver tolerance. 

To reiterate, the aim of this paper is to show that our framework learns the costate distribution of the fully parameterized problem, including inaccuracies that arise from the problem setup, control transcription, and numerical solver \cite{beeson2024globalsearchoptimalspacecraft}. 
Because the training data was generated with relatively coarse solver tolerances, the diffusion model samples match that distribution. 
When tighter tolerances are imposed, some samples fail to converge because the new parameterization defines a slightly different distribution. 
If the diffusion model was trained on data computed with a $10^{-6}$ NU tolerance, it should yield samples that converge more often under these tight tolerances. 
Generating data for training under the tighter tolerances will be more computationally expensive and our 
ongoing work investigates fine-tuning the diffusion model with a Markov Chain Monte Carlo algorithm, which may allow the generation of samples that satisfy tighter tolerances with limited new training data from the tight tolerance case.

\begin{table}[t!]
  \centering
  \caption{Percentage of diffusion model solutions for the GTO Halo transfer ($\alpha=0.95$) that remain feasible when warm-starting SNOPT under increasingly tighter feasibility tolerances with a maximum solving time of $30\,\mathrm{s}$. Each result is based on $1{,}000$ initial solutions for a feasibility tolerance of $10^{-3}$.}
  \label{tab: tol_refinement}
  \begingroup
  \setlength{\tabcolsep}{10pt}
  \renewcommand{\arraystretch}{1.1}
  \begin{tabular}{lcccc}
    \toprule
    \textbf{Feasibility Tolerance [NU]} & $10^{-3}$ & $10^{-4}$ & $10^{-5}$ & $10^{-6}$ \\
    \midrule
    \textbf{Convergence Rate [\%]} & 100.0 & 76.1 & 74.6 & 68.1 \\
    \bottomrule
  \end{tabular}
  \endgroup
\end{table}

\section{Conclusion}
This paper presents a machine learning framework that accelerates the time-intensive global search problem of low-thrust spacecraft trajectory optimization, with the novel contribution being the combination of generative models with the indirect approach for optimal control. 
Initial guesses for the costate variables are sampled from a diffusion model conditioned on the spacecraft's maximum thrust and used to warm-start a numerical optimizer. 
The strong performance of this approach is validated through its application to two transfer scenarios of varying complexity within the circular restricted three-body problem.
For training data generation, a global search strategy is implemented, which generates indirect control solutions at a high rate.
Complex structures in the costate space of feasible solutions for both problems are visualized using data reduction techniques. 
These global search structures vary with the spacecraft’s maximum thrust and are partially explained through the derivation of mathematical relations of the costates.
A diffusion model is  trained to learn this structure for fixed thrust levels in the training dataset and successfully predicts how it changes for new thrust levels not included in the training data.
The model’s performance in terms of generating a high number of solutions per minute for the testing thrust levels is evaluated by comparing it to samples from a uniform distribution within optimal ranges and samples from an adjoint control transformation.
Both methods are outperformed by the diffusion model by at least an order of magnitude for feasible solutions per minute.
For the Europa DRO problem, this improvement ranges from a factor of 20 to 75 depending on the thrust level, and for the GTO Halo transfer, from a factor of 11 to 26.

A main challenge of this method is the large amount of data required to train the model, leading to a big upfront cost for data generation.
Since the initial guesses sampled from the model are generally only as good as the data used to train it, this framework still needs to be combined with other methods which can generate high quality solutions. 
In future work, the framework will be further improved by generating training data at smaller intervals of the thrust level, with fewer data points per thrust level.
Additionally, the training cost will be reduced by training the model on principal components, which capture the majority of information in the data.
Furthermore, the application of the framework will be extended by conditioning it on different problem parameters and multiple parameters simultaneously.
Examples of such conditional problem parameters could be the boundary conditions, represented as energy levels of the initial or target orbit, the mass parameter $\mu$ of the CR3BP system or the objective value itself.
These extensions will further prepare the framework for potential application during preliminary mission design and, when combined with tighter tolerances and a higher-fidelity model, for potential autonomous on-board applications. 

\section{Acknowledgment}
The authors would like to thank Anjian Li for providing the foundational generative learning framework, which served as the basis for our approach, and informative discussions on diffusion modeling. 
Simulations were performed on computational resources managed and supported by Princeton Research Computing, a consortium of groups including the Princeton Institute for Computational Science and Engineering (PICSciE) and the Office of Information Technology’s High-Performance Computing Center and Visualization Laboratory at Princeton University.
The authors would also like to acknowledge partial support for this effort from a Princeton University School of Engineering and Applied Sciences internal Seed Grant award.

\section{Conflict of Interest}
On behalf of all authors, the corresponding author states that there is no conflict of interest.

\bibliographystyle{AAS_publication}   
\bibliography{references}   

\newpage
\section{Appendix}
\subsection{A. Network Architecture}
A U-Net consists of an encoder and a decoder. 
The encoder compresses the input data into a lower-dimensional hidden representation, resulting in a bottleneck layer. 
After the bottleneck, the decoder restores this representation to the original input dimension, with residual connections from the encoder to the decoder enhancing gradient flow.

The first layer in the U-Net is a convolutional layer that increases the channel size to the specified input dimension of the encoder block.
The encoder consists of three downsampling stages, where the channel number is increased and the sequence length is reduced at each stage. 
The channel number at each stage is specified as the encoder layer dimensions in Table \ref{tab:layer_sizes}.
Each stage of the downsampling process includes two ResNet blocks \cite{Kolesnikov.12242019}, group normalization \cite{Wu.3222018}, an attention block \cite{Vaswani.6122017}, and the downsampling operation.
The attention block plays a crucial role in the network, as it helps determine the strength of connections between elements in the input vector, which is essential for capturing relationships among the costates.

Since we use the U-Net to generate a vector instead of a two-dimensional image, a one-dimensional version of the original network used is employed. 
The input size is $(B\times C\times L)$, where $B$ is the batch size, $C$ is the channel number, and $L$ is the sequence length. 
Channel number and sequence length describe the dimension of the input data.
The batch size represents the subset of training data used during each step of the training process.

In addition to the feature matrix, each ResNet block requires information of the current timestep and the classifier as inputs. 
A sinusoidal positional embedding is applied to encode the current timestep $n$, increasing the dimensionality of the input by mapping the values to sine and cosine functions of different frequencies \cite{Vaswani.6122017}. 
The classifier information is encoded through two embedding class layers, each consisting of a linear layer combined with an activation function. 

The bottleneck layer consists of two more ResNet blocks, interleaved with another attention block. 
Following the bottleneck, three upsampling stages reverse the downsampling process by using the same building blocks, reducing the channel size while increasing the sequence length. 
After the upsampling stages, a final ResNet block and convolutional layer are applied, producing an output with the same dimensions as the input.
\newpage
\subsection*{B. Qualitative study of different diffusion model parameters}
\label{sec: appendix B}
\begin{figure}[h!]
    \centering
    \begin{subfigure}[b]{0.5\textwidth}
        \raggedright
        \includegraphics{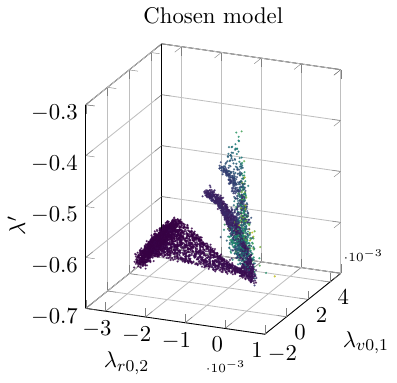}
    \end{subfigure}\hfill
    \begin{subfigure}[b]{0.5\textwidth}
        \raggedright
        \includegraphics{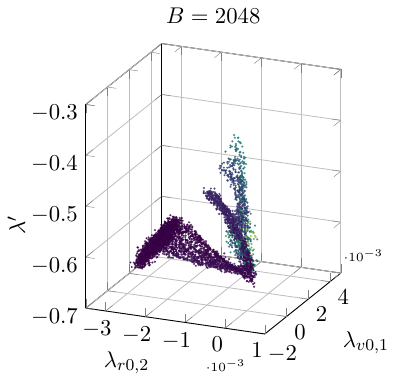}
    \end{subfigure}\hfill
    \begin{subfigure}[b]{0.5\textwidth}
        \raggedright
        \includegraphics{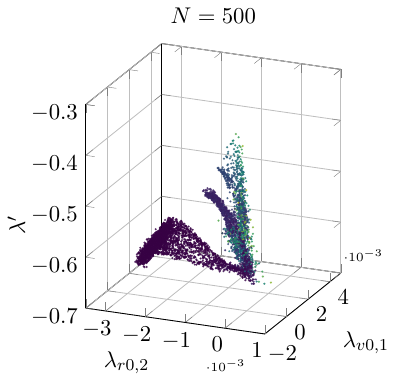}
    \end{subfigure}\hfill
    \begin{subfigure}[b]{0.5\textwidth}
        \raggedright
        \includegraphics{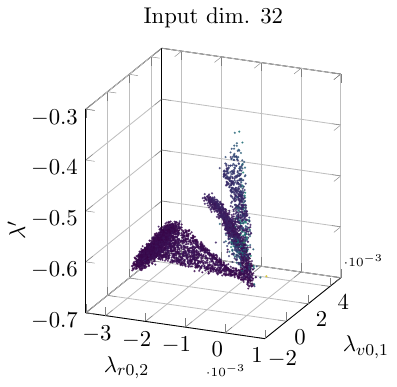}
    \end{subfigure}\hfill
    \begin{subfigure}[b]{0.5\textwidth}
        \raggedright
        \includegraphics{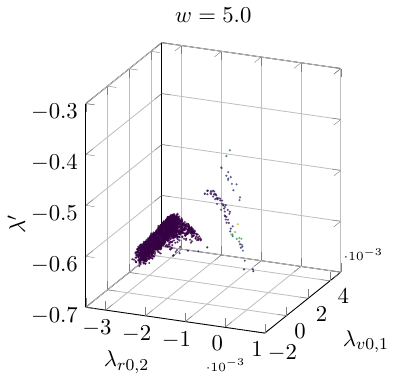}
    \end{subfigure}\hfill
    \begin{subfigure}[b]{0.5\textwidth}
        \raggedright
        \includegraphics{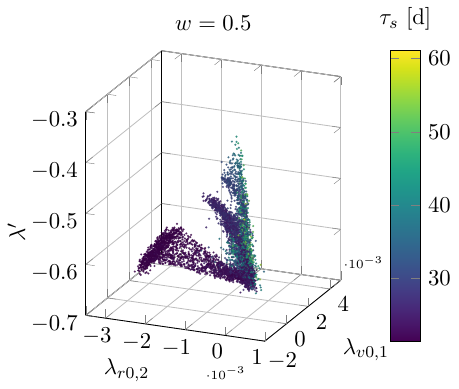}
    \end{subfigure}
    \caption{Structure in the costate space for diffusion-model samples with varying hyper-parameters ($\alpha=1.0$, Europa DRO transfer).  
    The reference model (parameters in Table~\ref{tab:layer_sizes}) is shown in the first panel; each remaining panel varies only the indicated parameter. Each plot depicts a subset of
    $5\,000$ samples.}
    \label{fig: model variations appendix}
\end{figure}

\end{document}